\documentclass[aps,prd,floatfix,superscriptaddress,preprintnumbers,floatfix]{revtex4} 
\usepackage{amssymb}
\usepackage{amsmath}
\usepackage{amsfonts}
\usepackage{epsfig}             
\usepackage{graphicx}
\usepackage{stackrel}
\usepackage{tabularx}
\usepackage{color}
\usepackage{dsfont}
\usepackage[T1]{fontenc}

\definecolor{darkerblue}{rgb}{0.0,0.0,0.5}
\newcommand{\seq}{\begin{subequations}}
\newcommand{\sen}{\end{subequations}}

\newcommand{\eq}{\begin{eqnarray}}
\newcommand{\en}{\end{eqnarray}}

\def\nn{\nonumber}

\begin{document}

\title{Test of vector portal with dark fermions in the charge-exchange reactions \\
  in the NA64 experiment at CERN SPS}

\author{Sergei N.~Gninenko}
\affiliation{Institute for Nuclear Research of the Russian Academy 
	of Sciences, 117312 Moscow, Russia} 

\author{Dmitry V.~Kirpichnikov}
\affiliation{Institute for Nuclear Research of the Russian Academy 
  of Sciences, 117312 Moscow, Russia}

\author{Sergey~Kuleshov} 
\affiliation{Millennium Institute for Subatomic Physics at 
the High-Energy Frontier (SAPHIR) of ANID, \\ 
	Fern\'andez Concha 700, Santiago, Chile}
\affiliation{Center for Theoretical and Experimental Particle Physics,
  Facultad de Ciencias Exactas, Universidad Andres Bello,
  Fernandez Concha 700, Santiago, Chile}
		
\author{Valery~E.~Lyubovitskij} 
\affiliation{Departamento de F\'\i sica y Centro Cient\'\i fico
Tecnol\'ogico de Valpara\'\i so-CCTVal, \\ 
Universidad T\'ecnica Federico Santa Mar\'\i a,
Casilla 110-V, Valpara\'\i so, Chile}
\affiliation{Millennium Institute for Subatomic Physics at
the High-Energy Frontier (SAPHIR) of ANID, \\
Fern\'andez Concha 700, Santiago, Chile}

\author{Alexey~S.~Zhevlakov} 
\affiliation{Bogoliubov Laboratory of Theoretical Physics, JINR,
 141980 Dubna, Russia} 
\affiliation{Matrosov Institute for System Dynamics and 
 Control Theory SB RAS, \\  Lermontov str., 134,
 664033, Irkutsk, Russia } 

\begin{abstract}

We consider an experiment to search for dark sector particles in dark photon kinetic mixing model 
by analyze invisible and semi-invisible decays of neutral mesons  $M^0 = \pi^0$, $\eta$, $\eta'$,
$\omega$, $f_2(1270)$, produced in the NA64 experiment at the CERN SPS. The approach proposed in Ref.~\cite{Gninenko:2014sxa} is to use 
the charge-exchange reactions $\pi^- + (A, Z) \to M^0 + (A,Z-1); M^0 \to$ {\it invisible} or {\it semi-invisible} of high-energy pions (or kaons) at a nuclei target as a source of $M^0$s, 
which subsequently decay invisibly into dark sector.  This reaction chain would lead to a striking signature of the signal event - the complete disappearance of the beam energy in the setup. 
Using data obtained from the  study of charge-exchange reactions
at IHEP (Protvino) and Fermilab (Batavia) we show that the 
integral cross sections $\sigma$ for production of 
the neutral mesons $M^0$ are slightly deviate from 
phenomenological formula $\sigma \sim Z^{2/3}$, where $Z$ is
the nuclei charge. 
In particular, we present the formulas for the differential and integral sections that explicitly depend on the Mandelstam and $Z$ variables. Derived formulas are used to predict the cross sections as a function of beam energy  for several target nuclei, and to estimate the projection sensitivity for the proposed search for the $M^0\to$ {\it semi-invisible} and $M^0\to$ {\it invisible} decays through the vector portal to dark sector. Sensitivity to different semi-invisible decay modes of neutral pseudoscalar mesons is studied. 
\end{abstract}
		
\maketitle
	
\section{Introduction}
Nowadays searches for Dark Matter (DM) and study of its properties are hot challenges in particle 
physics, astrophysics, and cosmology. In the last decade,  a broad experimental program has been developed to detect
 non-gravitational DM interactions, including  direct searches for  DM by measuring 
 recoil energy from DM-nucleus scattering, indirect searches for particles  from DM annihilation, 
 and accelerator-based measurements \cite{Essig:2013lka,Alexander:2016aln, Battaglieri:2017aum, Beacham:2019nyx, Agrawal:2021dbo}.
 For the thermal DM in sub-GeV mass range,  we have to assume the existence of a new feeble interaction between the ordinary and dark matter.  One can stress several hidden sector scenarios that have been widely discussed in literature when such interaction is transmitted through the  
Higgs~\cite{Arcadi:2019lka,Davoudiasl:2021mjy}, tensor~\cite{Voronchikhin:2023znz,Voronchikhin:2022rwc,Kang:2020huh}, 
vector or dark photon, see, e.g.~\cite{Bauer:2018onh, Fabbrichesi:2020wbt, Lanfranchi:2020crw,Buras:2021btx,Kachanovich:2021eqa,Lyubovitskij:2022hna},  
sterile neutrino~\cite{Escudero:2016tzx},  and axion or axion-like (ALPs)~\cite{Nomura:2008ru,Zhevlakov:2022vio,Armando:2023zwz} portals.
Studies at accelerators have mostly focused on testing of models  in which DM couples to the SM through  a new massive gauge boson (a dark photon $A'$) that kinetically mixes with the ordinary 
photon~\cite{Okun:1982xi, Holdom:1985ag} which interacts universally with lepton and quarks. Such experiments searched for  light sub-GeV DM by looking either for visible or invisible decays of  the $A'$, or another mediator. 
The searches for invisible decays  typically use missing energy (momentum)  techniques developed for leptonic ($e$ or $\mu$)  
beams~\cite{Gninenko:2013rka,Gninenko:2014pea,Izaguirre:2014bca}.  
 
 \par   Recently, significant attention has been received by alternative models with light DM scenarios when the interaction between  
  DM and the SM is primarily leptophobic (or hadrophilic), i.e. the corresponding mediator couples  predominantly to quarks, see, e.g.  Refs.~\cite{Tulin:2014tya,Dobrescu:2014fca,Soper:2014ska}.
  Taking into account the low  level of our knowledge about the nature of DM, such scenarios are certainly worth studying. This also emphasizes the need for a broad experimental program using both hadron and lepton beams.
   One   possibility to probe leptophobic  dark sector is to use neutrino or beam dump experiments \cite{Dobrescu:2014ita, Batell:2014yra, Batell:2018fqo,COHERENT:2022pli}.
  Another possibility is to look for invisible  decays of neutral mesons into dark sector particles at  $e^+e^-$ colliders 
  \cite{BES:2006pmv,BESIII:2012nen}.   Experimental studies of invisible hadronic decays were performed by several collaborations. In 
particular, the BES III collaboration~\cite{BESIII:2012nen,BESIII:2018bec} set the constraints on the invisible 
branching fraction of the $\eta$, $\eta'$, $\omega$, and $\phi$ mesons.  
The BABAR Collaboration~\cite{BaBar:2009gco,BaBar:2013npw} has studied the invisible decay modes of heavy quarkonia. The NA62 Collaboration~\cite{NA62:2020pwi} established the limits on invisible decays of $\pi^0$. 
Future experiments plan to collect sufficiently large projected 
statistics of mesons. In this vein, we would like to mention the experiments: $K$-meson factory (HIKE 
experiment~\cite{HIKE:2022qra}), $\eta$/$\eta'$ factories (REDTOP 
experiment~\cite{REDTOP:2022slw,Gatto:2019dhj} and HIAF experiment~\cite{HIAF}), and the Forward Physics Facility at CERN \cite{Batell:2021snh}.     

\par In this work we discuss    the  experiment,   proposed in Ref. \cite{Gninenko:2014sxa},   to search for leptophobic dark sector particles    in the  decays of neutral mesons  $M^0 = \pi^0$, $\eta$, $\eta'$, $K^0_{L,S}$,...
produced in the NA64 experiment at the CERN SPS. The idea is to use  the   charge-exchange reactions $\pi^- + (A, Z) \to M^0 + (A,Z-1); M^0\to invisible$  of high-energy pions (or kaons) at a  nuclei target  $(A,Z)$ as a source of  $M^0$s with their subsequent invisible decay into dark sector particles. 
 The great advantage of using $M^0 \to invisible$ decays is that their invisible decay rate into a couple of neutrinos in the SM  
is highly suppressed in the SM, $\Gamma(M^0 \to \nu \nu)/\Gamma(M^0 \to total) \ll 10^{-16}$ for masses $m_\nu \simeq 10$ eV and $m_{M^0} \simeq m_{K^0}$.  We 
show that this  rate of $M^0 \to invisible$ or  $M^0 \to semi-invisible$ decays could be enhanced up to a measurable level due to the contribution from the dark sector physics. 
The striking signature of this reaction chain is the complete disappearance  of the primary beam energy in the setup which  we intend to probe at hadronic  beams in NA64  with missing energy technique.  This approach logically complements  the use of leptonic beams in the scientific program of the NA64 experiment at the CERN SPS for further constraining DM parameter space. In particular, by using SPS  facilities to accumulate high statistics of particles on target the NA64 experiment can use charged pion and kaon beams as a  base for the widening dark sector program at NA64 with testing  DM physics in hadronic processes (NA64h). During the last two 2022 and 2023 runs within a few days of data taking,  NA64 accumulated $ \simeq 10^{10}$ 
pions on target ($\pi$OT) in order to understand the potential of the experiment in the leptophobic dark sector by using the pion beam and missing energy technique. During 3-4 months of running, collection of $\sim 10^{12}$  $\pi^-$OT seems possible. 

\par Below  we will estimate bounds on parameter spaces of the vector portal with dark fermions. 
The first theoretical analysis of parameter space of DM based on using invisible neutral kaon decays was performed in Refs. \cite{Gninenko:2015mea,Gninenko:2016rjm, Barducci:2018rlx, Hostert:2020xku, Hostert:2020gou}, and for invisible 
vector meson decays  was done in Ref.~\cite{Schuster:2021mlr}. In particular, in Ref.~\cite{Schuster:2021mlr} 
it was estimated a yield of vector mesons with a  lepton beam scattered at a heavy nuclei target. 
In Ref.~\cite{Zhevlakov:2023wel} a novel idea to search for invisible 
decay of the vector $\rho^0$ meson produced at the accelerators with hadronic beams 
have been further developed in context of the NA64h  experimental setup by   
using cut to small recoil energy at meson production. 
In the present work we will use a model-independent estimate for yield of mesons, which is based on 
experimental data of diffraction processes $\pi^- + (A, Z) \to M^0 + (A,Z-1)$ in the Regge regime. The data have been collected in a series of 
experiments~\cite{Barnes:1976ek,Dahl:1976em,Serpukhov-CERN:1979mvf,
Serpukhov-CERN:1978unh,Serpukhov-CERN:1979vql,
Apokin:1981iv,Apokin:1986ka,Donskov:2013uva}. Missing 
energy technique gives a possibility to establish conservative limits of invisible or semi-visible neutral 
meson decay after creation at scattering charged pion beam at a fixed target.                 
The paper is organized as follows. In Sec.~\ref{framework} we briefly discuss missing energy conception to 
analyze invisible and semi-invisible neutral meson decay modes for 
the case of hadronic beam in the NA64 experiment. In Sec.~\ref{RechargeSec}  we present a model-independent estimate for the charge exchange process for different $Z$ charges of atomic targets. We also  discuss invisible
and  semi-invisible decays of neutral mesons in Sec.~\ref{Sec:decays}. The results from invisible and semi-
invisible neutral meson decay mode to DM fermions and implementation to DM parameter space are presented in 
Sec.~\ref{bounds_sec}.  Finally, in Sec.~\ref{Conclusion} we give our conclusions.

\section{Framework}
\label{framework}

In addition to the direct search for DM, and using existing experiments of $e^-e^+$ collision
or meson decay in flight, was proposed a conception to search for DM signals in fixed target experiments by 
measurement of missing energy or momenta technique~\cite{Gninenko:2013rka}. 
It was realized at NA64 at CERN SPS experiments with electron~\cite{NA64:2023wbi}, 
muon~\cite{Sieber:2021fue}, and positron~\cite{NA64:2023ehh} beams.  
 In the future, it is planned to run several new experiments to study sub-GeV DM region, e.g., 
M$^3$~\cite{Kahn:2018cqs,Capdevilla:2021kcf},
LDMX~\cite{Berlin:2018bsc,LDMX:2018cma,Ankowski:2019mfd,Schuster:2021mlr,Akesson:2022vza}, 
BDX~\cite{BDX:2014pkr,Izaguirre:2014dua,BDX:2016akw}, 
and ILC~\cite{Asai:2023dzs, ILC:2013jhg}. 

In order to use hadronic mode in the NA64 experiment for direct   
 DM production one needs to formulate a strategy. 
Search for DM by studying the invisible decay widths of neutral mesons  is one of the usable instruments. It is 
required to estimate the yield of neutral mesons at charged pion scattering onto a fixed target.  In addition, the experimental facility is required to measure a small recoil energy which can be a source of the background in the detector. In our case,  we deal with the diffraction process,  where the main contribution to background is negligible 
at small  energy transfer in $t$- channel. In next section we will discuss results for 
the integral cross sections for the charge-exchange reactions $\pi^- + (A, Z) \to M^0 + (A,Z-1)$ for different types of nuclear targets and produced mesons $M^0$. 

The yield of neutral mesons $M^0$ produced by $\pi^-$ beam scattering at a fixed target is
\begin{equation}
N_{M^0} \simeq \pi\mbox{OT}\cdot \frac{\rho_{T} N_A}{A} L_T \,\, 
\sigma_{2\to2}(\pi^- + (A, Z) \to M^0 + (A,Z-1)) \,, 
\label{Nrho}
\end{equation}
where $A$ is the atomic weight number, $N_A$ is the Avogadro's number, $\pi\mbox{OT}$ is the number of 
negative charged pions accumulated on target, $\rho_{T}$ is the target density, 
$L_T$ is the effective thickness of the target which in conservative scenario is assumed 
to be equal to effective pion interaction length in the target~\cite{Workman:2022ynf},
$\sigma_{2\to2}(\pi^- + (A, Z) \to M^0 + (A,Z-1))$ is the cross section of charge-exchange reaction.  The value of the cross section for the charge-exchange reaction will be considered in the next section.

In our estimate,  we consider a modified experimental setup of the NA64 with a negative charged pion beam
scattered at the active iron target (NA64$_h$ setup). The experiment NA64$_h$ also employs  
a hadronic calorimeter which represents four or three modules in 48 layers (2.5 mm of iron plates and 
4 mm of scintillator). 
We estimate limit on the invisible branching ratio 
of produced neutral mesons, $M^0$,  at 90\% confidence level (C.L.) assuming zero observed signal events and background-free case, which implies  $\mathrm{Br}(M^0\to \text{inv.})\leq 2.3/N_{M^0}$ for invisible and for 
semi-invisible is $\mathrm{Br}(M^0\to \text{semi-inv.})\leq 2.3/N_{M^0}$, where $N_{M^0}$ is a number of the produced vector mesons from Eq.~(\ref{Nrho}).   
About $3\times 10^{9}$ pions on target $(\pi \mbox{OT})$ in the NA64 experiment were accumulated in a short period  of technical data taking at CERN SPS. 
For the typical projected statistics of NA64$_h$ we will use the number of $\sim 10^{12} \pi\mbox{OT}$ which can be accumulated during 1 year of the experimental run. The optimistic projected collected data limit which we will use here is $5\times 10^{12} \pi\mbox{OT}$.

\section{Charge-exchange reactions}
\label{RechargeSec}

\begin{figure}[t]
 \includegraphics[scale=.65]{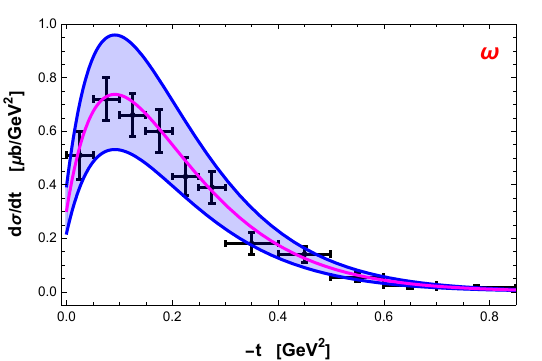}
  \includegraphics[scale=.65]{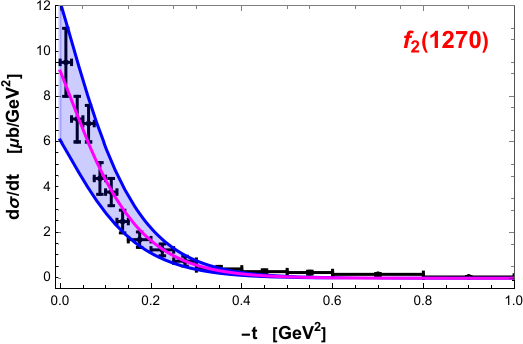}
    \includegraphics[scale=.65]{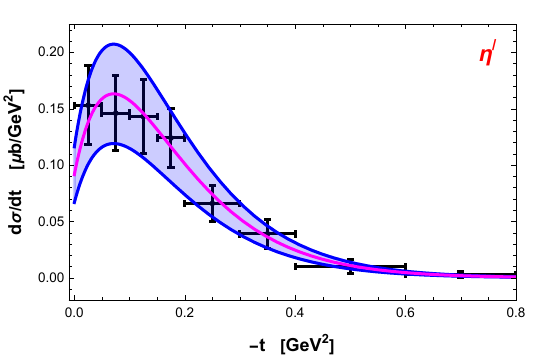}
    \caption{Fit of the parameters for the 
    $\omega$, $f_2(1270)$, and $\eta^\prime$ mesons
using IHEP data for the differential cross section of their production at $P=39.1$ GeV beam~\cite{Apokin:1986ka}. }
\label{fig:1}
\end{figure}

Charge-exchange reactions $\pi^- + (A, Z) \to M^0 + (A,Z-1)$
at nuclear target $(A,Z)$ with $M^0 = \pi^0$, $\eta$, $\eta'$,
$\omega$, $f_2(1270)$~\cite{Barnes:1976ek,Dahl:1976em,Serpukhov-CERN:1979mvf,
Serpukhov-CERN:1978unh,Serpukhov-CERN:1979vql,
Apokin:1981iv,Apokin:1986ka,Donskov:2013uva} 
give a unique possibility to shed light on hadron structure,
Regge phenomenology~\cite{Gribov:2009zz,Collins:1977jy}, 
and color transparency~\cite{Kopeliovich:1991wu}. 
In particular,
the amplitude ${\cal M}(s,t)$ of the charge-exchange reaction depending
on the $s$ and $t$ Mandelstam variables, can be factorized
as~\cite{Gribov:2009zz,Collins:1977jy}: ${\cal M}(s,t) \sim A(t)
(s/s_0)^{\alpha_r(t)}$. Here $A(t)$ is a phenomenological
function fitted from data, $s_0 = 10$ GeV$^2$ is the input
total energy for the evolution of the cross section,
$\alpha_r(t)$ is the Regge trajectory.
$\alpha_r(t)$ is normally straight line parameterized as
$\alpha_r(t) = \alpha(0) + \alpha' t$. In case of small
$|t| \ll s$ one can neglect by the $t$ dependence
in the Regge trajectory. Indeed, all data on the 
charge-exchange reactions are well described with constant
Regge trajectory $\alpha_r(0) \simeq \alpha(0)  \sim  0.5$.
Differential cross section in case of the proton target
is parameterized as~\cite{Serpukhov-CERN:1979vql,Serpukhov-CERN:1978unh,Serpukhov-CERN:1979mvf}:  
\eq\label{dsigma1} 
\frac{d\sigma_H(s,t)}{dt} = \frac{d\sigma_H(s,t)}{dt}\bigg|_{t=0} 
\  \biggl[1 - g(s) c(s) t\biggr] \ 
\exp[c(s) t] \,, 
\en
where 
\eq\label{dsigma2} 
\frac{d\sigma_H(s,t)}{dt}\bigg|_{t=0} = A
\, \Big(\frac{s}{s_0}\Big)^{2 \alpha_r(0) - 2} \,, 
\en
$A$ is the normalization factor, 
  $g(s) = g_0 + g_1 \log(s/s_0)$ and
  $c(s) = c_0 + c_1 \log(s/s_0)$ are the $s$-running couplings. 
  For the specific reaction with
  $\pi^0$, $\eta$, $\eta'$, $\omega$, and $f_2(1270)$ 
  production the sets of parameters
  are shown in Table~\ref{tab:1} (some of them have been fixed in
  Refs.~\cite{Serpukhov-CERN:1979vql,Serpukhov-CERN:1978unh,%
  Serpukhov-CERN:1979mvf}. For proton/neutron target these processes 
  were studied in Ref.~\cite{Nys:2018vck}.  In Fig.~\ref{fig:1} we present the results of the fit of the parameters 
defining the parameterizations for the differential cross sections 
of the charge-exchange reactions on the proton target: 
$\pi^-  + p \to M^0 + n$ 
for the cases of the $M^0 = \omega$, $f_2(1270)$, and $\eta^\prime$ meson 
production using IHEP data for the $d\sigma_H(s,t)/dt$ at $P=39.1$ GeV beam~\cite{Apokin:1986ka}. 
 
For the integral cross section in case of the  proton target we get 
\eq\label{sigma_P}
\sigma_H(s) = 
A \, \Big(\frac{s}{s_0}\Big)^{2 \alpha_r(0) - 2} \,
\, \frac{1 + g(s)}{c(s)} \,.  
\en

Extension to arbitrary nuclei $N$ with charge $Z$ is normally
done by multiplying with factor $Z^{2/3}$. However, we found
that this behavior should be slightly corrected as
$Z^{2/3-0.15 Z^{-2/3}}$. I.e. the integral cross section
for the neutral meson production
at nuclei with charge $Z$ reads:
\eq\label{sigma_Z}
\sigma_N(s) = \sigma_H(s) \, Z^{2/3-0.15 Z^{-2/3}} 
\,. 
\en
E.g., our prediction for the ratio of cross sections
of production on different nuclei does not depend on the
total energy $s$ and produced pseudoscalar meson,
while depends on $Z$:
$\sigma_N(s)/\sigma_H(s) = Z^{2/3-0.15 Z^{-2/3}}$. E.g.,
the ratio of the total cross section of productions
on carbon and hydrogen is 3.04, which is in good
agreement with data: $3.2 \pm 0.1$~\cite{Apokin:1986ka}.   

\begin{figure}[h]
\begin{center}
  \includegraphics[scale=.55]{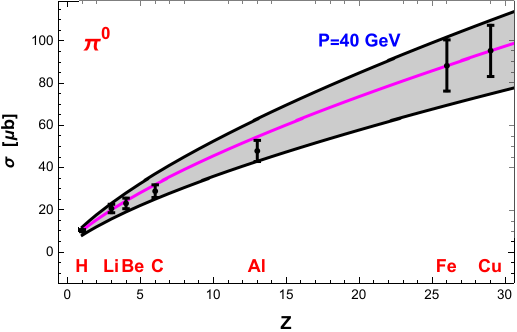}
  \hspace*{.2cm}
  \includegraphics[scale=.55]{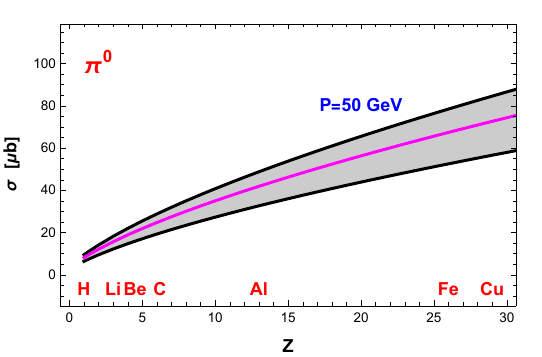}
\end{center}
\caption{Results for the integral cross section of the $\pi^0$ production  $\pi^-  + (A,Z) \to \pi^0 (\to 2  \gamma) + (A,Z-1)$:
comparison of parameterization~(\ref{sigma_P}) and~(\ref{sigma_Z}) 
with data for the beam momentum $P=40$ GeV~\cite{Apokin:1981iv} 
(left panel), predictions for $P=50$ GeV (right panel). 
    \label{fig:2}}
\end{figure}

\begin{figure}[h]
\begin{center}
\includegraphics[scale=.55]{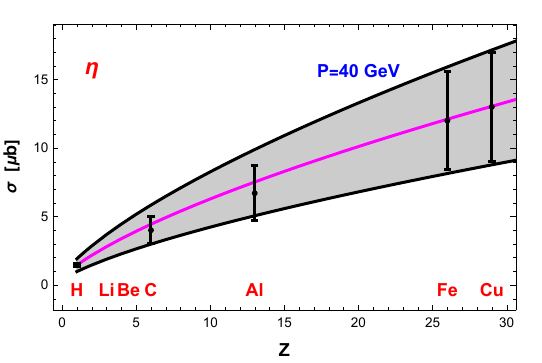}
\includegraphics[scale=.55]{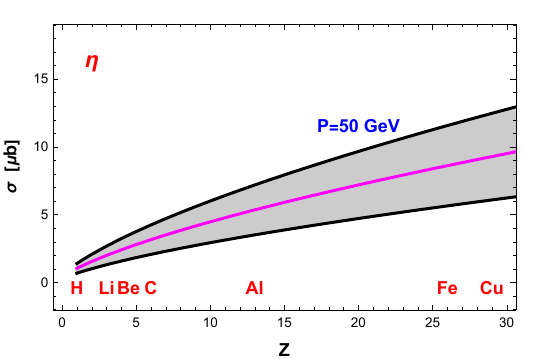}
\end{center}
\vspace*{-.5cm}
\noindent
\caption{Results for the integral cross section of the 
$\eta$ production $\pi^-  + (A,Z) \to \eta (\to 2  \gamma) + (A,Z-1)$:
comparison of parameterization~(\ref{sigma_P}) and~(\ref{sigma_Z}) 
with data for the beam momentum $P=40$ GeV~\cite{Apokin:1981iv} 
(left panel), predictions for $P=50$ GeV (right panel). 
  \label{fig:3}}
\end{figure}

\begin{figure}[h]
\begin{center}
\includegraphics[scale=.55]{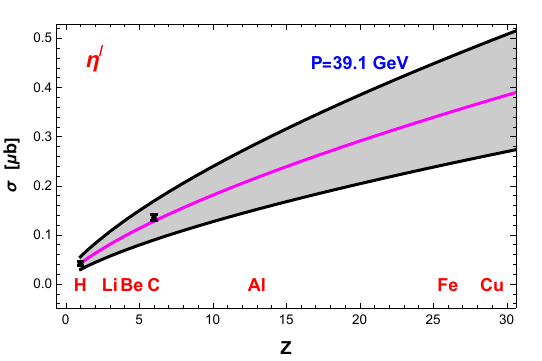}
\includegraphics[scale=.55]{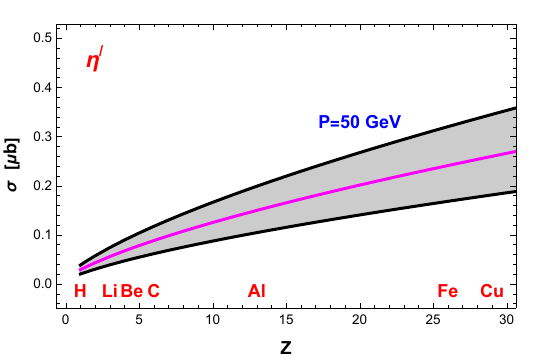}
\end{center}
\vspace*{-.5cm}
\noindent
\caption{Results for the integral cross section of the 
$\eta'$ production $\pi^-  + (A,Z) \to \eta' (\to 2  \gamma) + (A,Z-1)$: 
comparison of parameterization~(\ref{sigma_P}) and~(\ref{sigma_Z}) 
with data for the beam momentum $P=39.1$ GeV~\cite{Apokin:1986ka} 
(left panel), predictions for $P=50$ GeV (right panel). 
  \label{fig:4}}
 \end{figure}

\begin{figure}
\begin{center}
\includegraphics[scale=.55]{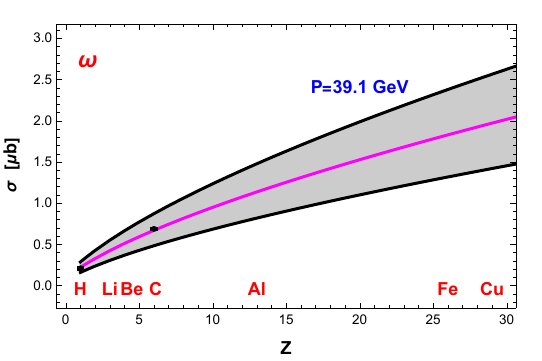}
\includegraphics[scale=.55]{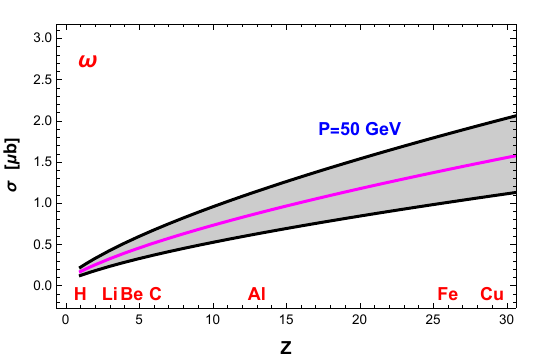}
\end{center}
\vspace*{-.5cm}
\noindent
\caption{Results for the integral cross section of the 
$\omega$ production 
$\pi^-  + (A,Z) \to \omega (\to \pi^0 \gamma) + (A,Z-1)$: 
comparison of parameterization~(\ref{sigma_P}) and~(\ref{sigma_Z}) 
with data for the beam momentum $P=39.1$ GeV~\cite{Apokin:1986ka} 
(left panel), predictions for $P=50$ GeV (right panel). 
\label{fig:5}}
\end{figure}

\begin{figure}
\begin{center}
\includegraphics[scale=.55]{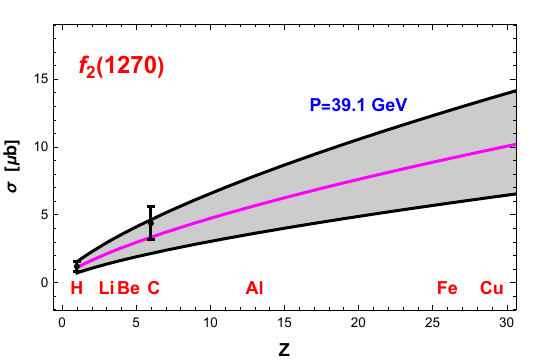}
\includegraphics[scale=.55]{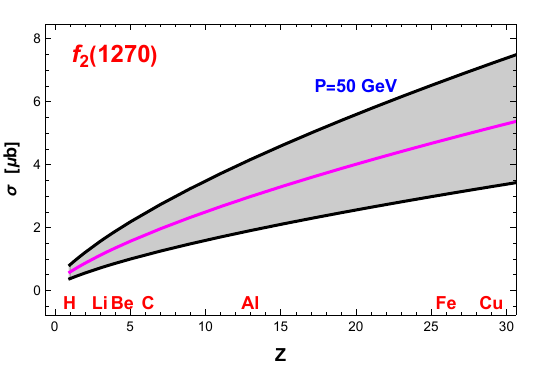}
\end{center}
\vspace*{-.5cm}
\noindent 
\caption{Results for the integral cross section of the 
$f_2(1270)$ production 
$\pi^-  + (A,Z) \to f_2(1270) (\to 2  \pi^0) + (A,Z-1)$: 
comparison of parameterization~(\ref{sigma_P}) and~(\ref{sigma_Z}) 
with data for the beam momentum $P=39.1$ GeV~\cite{Apokin:1986ka} 
(left panel), predictions for $P=50$ GeV (right panel). 
\label{fig:6}}
\end{figure}

In Figs.~\ref{fig:2}-\ref{fig:6} for each type of the meson 
($\pi^0$, $\eta$, $\eta'$, $\omega$, $f_2(1270)$) we present two plots 
for the integral cross sections of the charge-exchange reactions 
$\pi^-  + (A,Z) \to M^0 (\to 2  \gamma) + (A,Z-1)$ with $M^0 = \pi^0, \eta, \eta'$, 
$\pi^-  + (A,Z) \to \omega (\to \pi^0  \gamma) + (A,Z-1)$, and 
$\pi^-  + (A,Z) \to  f_2(1270) (\to 2 \pi^0) + (A,Z-1)$ 
on the different nuclear targets including ${\rm H}$, 
${\rm Li}$, ${\rm Be}$, ${\rm C}$, ${\rm Al}$, ${\rm Fe}$, and 
${\rm Cu}$. In the left panel we present a comparison of the 
parametrization~(\ref{sigma_P}) and~(\ref{sigma_Z}) with data  
for beam momentum $P=40$ GeV or for $P=39.1$~GeV. In the right 
panel we present our predictions for the $P=50$~GeV. 
For the case of ${\rm Fe}$ target we made the predictions 
using the $Z$ dependence of the integral cross section established in 
Eq.~(\ref{sigma_Z}). 

In Tables~\ref{tab:2} and~\ref{tab:3} 
we display our predictions for the integral cross 
sections for the reactions $\pi^-  + (A,Z) \to M^0 + (A,Z-1)$, 
$\pi^-  + (A,Z) \to  P^0 (\to 2  \gamma) + (A,Z-1)$, 
$\pi^-  + (A,Z) \to \omega (\to \pi^0  \gamma) + (A,Z-1)$, 
and $\pi^-  + (A,Z) \to  f_2(1270) (\to 2 \pi^0) + (A,Z-1)$ 
at $P=50$ GeV. Here $P^0 = \pi^0, \eta, \eta'$. 
We take the central values of the branchings of the neutral mesons
from Particle Data Group~\cite{Workman:2022ynf}:
\eq
{\rm Br}(\pi^0 \to 2 \gamma) &=& 0.99\,,
\nonumber\\
{\rm Br}(\eta \to 2 \gamma) &=& 0.3936\,, 
\nonumber\\
{\rm Br}(\eta' \to 2 \gamma) &=& 0.02307\,,
\nonumber\\
{\rm Br}(\omega \to \pi^0 \gamma) &=& 0.0835\,,
\nonumber\\
{\rm Br}(f_2 \to 2 \pi^0) &=& 0.281\,.     
\en 

One can see good agreement of the parameterization~(\ref{sigma_P}) and~(\ref{sigma_Z}) with data. 
 However, we would like to note that for mesons with masses $m_{M^0} > m_{\pi^0}$  the current precision for  their  total production cross section 
is at the level $\simeq 30-35\%$, see Tables \ref{tab:2} and \ref{tab:3}.  
This will constraints the sensitivity of the future searches 
for $M^0\to invisible$ decays, and thus, this precision  should be improved by 
at least a factor of $\gtrsim$ 3.   

\begin{table}[ht]
\begin{center}
\caption{Parameters of the differential cross sections}

\vspace*{.1cm}

\def\arraystretch{1.25}
\begin{tabular}{|c|c|c|c|c|c|c|}
\hline
Meson & $A$ & $\alpha_r$
& $c_0$ (GeV$^{-2}$) & $c_1$ (GeV$^{-2}$)
& $g_0$ & $g_1$  \\
\hline
$\pi^0$ & $430 \pm 20$ & $0.48 \pm 0.01$ 
& $12.7 \pm 0.3$ & $1.57 \pm  0.12$
& $2.55 \pm 0.09$ & $- 0.23 \pm 0.06$ \\
\hline
$\eta$  & $36 \pm 2$ & $0.37 \pm 0.02$ 
& $6 \pm 0.2$ & $1.60 \pm 0.10$ & 
$4.6 \pm 0.3$ & - $0.5 \pm 0.2$ \\
\hline
$\eta'$ & $1.37 \pm 0.37$ & $0.325 \pm 0.01$ & 6.84 & 1.7 & 3.7 & 0 \\
\hline
$\omega$ & $2 \pm 0.5$ & $0.53 \pm 0.01$ & 6.5 & 1.23 & 5.5 & 0 \\
\hline
$f_2$  & $60  \pm 20$   & $0.53  \pm 0.01$  & 8 & 2.6 & 4.60 & - 2 \\
\hline
\end{tabular}
\label{tab:1}
\end{center}

\begin{center}
  \caption{Predictions for the integral cross
    sections for the $\pi^-  + (A,Z) \to M^0 + (A,Z-1)$
    reactions
    at beam momentum $P=50$ GeV in $\mu$b units.}

\vspace*{.1cm}

\def\arraystretch{1.25}
\begin{tabular}{|c|c|c|c|c|c|c|c|}
\hline
Meson & H & Li & Be & C & Al & Fe & Cu \\
\hline
$\pi^0$
& $ 8.1 \pm 1.4$
& $15.6 \pm 2.6$
& $18.8 \pm 3.1$
& $24.7 \pm 4.1$
& $41.9 \pm 6.8$
& $67.4 \pm 11$
& $72.6 \pm 11$
\\
\hline
$\eta$
& $2.6 \pm 0.9$
& $5.1 \pm 1.7$
& $6.1 \pm 2.1$
& $8.0 \pm 2.8$
& $13.6 \pm 4.7$
& $21.9 \pm 7.5$
& $23.6 \pm 8.1$
\\
\hline
$\eta'$
& $1.3 \pm 0.4$
& $2.4 \pm 0.8$
& $2.9 \pm 1.0$
& $3.8 \pm 1.3$
& $6.5 \pm 2.1$
& $10.4 \pm 3.5$
& $11.3 \pm 3.7$
\\
\hline
$\omega$
& $2.0 \pm 0.7$
& $3.9 \pm 1.2$
& $4.7 \pm 1.5$
& $6.2 \pm 1.9$
& $10.5 \pm 3.2$
& $16.9 \pm 5.1$
& $18.2 \pm 5.6$
\\
\hline
$f_2$
& $2.1 \pm 0.8$
& $4.0 \pm 1.5$
& $4.8 \pm 1.9$
& $6.3 \pm 2.4$
& $10.6 \pm 4.2$
& $17.1 \pm 6.7$
& $18.4 \pm 7.3$
\\
\hline
\end{tabular}
\label{tab:2}
\end{center}

\vspace*{.5cm}

\begin{center}
  \caption{Predictions for the integral cross
    sections for the
    $\pi^-  + (A,Z) \to  P^0 
    (\to 2  \gamma) + (A,Z-1)$,
    $\pi^-  + (A,Z) \to \omega
    (\to \pi^0  \gamma) + (A,Z-1)$,
    and $\pi^-  + (A,Z) \to  f_2 (\to 2 \pi^0) + (A,Z-1)$,
    reactions
    at beam momentum $P=50$ GeV in $\mu$b units.
    Here $P^0 = \pi^0,\eta,\eta'$.}


\def\arraystretch{1.25}
\begin{tabular}{|c|c|c|c|c|c|c|c|}
\hline
Meson & H & Li & Be & C & Al & Fe & Cu \\
\hline
\!\!
$\pi^0$ 
& $ 8.1 \pm 1.4$
& $15.6 \pm 2.6$
& $18.8 \pm 3.1$
& $24.7 \pm 4.1$
& $41.9 \pm 6.8$
& $67.4 \pm 11$
& $72.6 \pm 11$
\\
\hline
$\eta$
& $1.0 \pm 0.4$
& $2.0 \pm 0.7$
& $2.4 \pm 0.8$
& $3.2 \pm 1.0$
& $5.4 \pm 1.8$
& $8.6 \pm 3.0$
& $9.3 \pm 3.2$
\\
\hline
$\eta'$
& $0.03 \pm 0.01$
& $0.06 \pm 0.01$
& $0.07 \pm 0.03$
& $0.09 \pm 0.03$
& $0.15 \pm 0.05$
& $0.24 \pm 0.08$
& $0.26 \pm 0.09$
\\
\hline
$\omega$
& $0.17 \pm 0.05$
& $0.33 \pm 0.10$
& $0.39 \pm 0.12$
& $0.52 \pm 0.15$
& $0.87 \pm 0.27$
& $1.41 \pm 0.43$
& $1.52 \pm 0.46$
\\
\hline
$f_2$
& $0.58 \pm 0.23$
& $1.11 \pm 0.44$
& $1.34 \pm 0.53$
& $1.76 \pm 0.69$
& $2.98 \pm 1.18$
& $4.80 \pm 1.89$
& $5.17 \pm 2.04$
\\
\hline
\end{tabular}
\label{tab:3}
\end{center}
\end{table}

In Fig.~\ref{fig:1} we present the results of the fit of the parameters defining the parameterizations for the differential cross sections at proton target for the cases of the $\omega$, $f_2(1270)$, and $\eta^\prime$ meson production 
using IHEP data for the $d\sigma_H(s,t)/dt$ at $P=39.1$ GeV beam~\cite{Apokin:1986ka}. 
In Figs.~\ref{fig:2}-\ref{fig:6} for each type of the meson 
($\pi^0$, $\eta$, $\eta'$, $\omega$, $f_2(1270)$) 
we present two plots: 
comparison of the results for beam momentum $P=40$ GeV 
or for $P=39.1$ GeV obtained using formulas~(\ref{sigma_P}) and~(\ref{sigma_Z})
with data~\cite{Apokin:1981iv,Apokin:1986ka} (left panel)
and predictions for the $P=50$ GeV. In case of the Fe target we 
made the predictions using the formula for the integral cross section derived for arbitrary $Z$ (right panel).
One can see good agreement of parameterization~(\ref{sigma_P}) and~(\ref{sigma_Z}) 
with data. 


\clearpage

\section{Invisible neutral meson decays to dark fermions \\
through dark photon portal}
\label{Sec:decays}
   
 In this section,  we will discuss invisible neutral meson decays to dark fermions through a dark photon portal. The dark photon portal 
can be introduced via kinetic mixing with 
the SM photons~\cite{Holdom:1985ag}. In particular, 
the gauge invariant coupling of the dark photon $A'$ and the SM photon $A$ has the form 
\eq\label{Lmix} 
\mathcal{L}_{mix} = \frac{\epsilon}{2} F_{\mu\nu}A'^{\mu\nu}
\en
where $\epsilon$ is the mixing parameter, $F_{\mu\nu}$ and $A'_{\mu\nu}$ are the stress tensors of the $A$ and $A'$ fields, respectively. 
We would like to mention that the derivation of the dark photon via the Stueckelberg mechanism was considered in Ref.~\cite{Kachanovich:2021eqa}.

The interaction of the dark photon with the charged current of SM fermions and with dark fermion current has a form:  
\eq
\mathcal{L}\supset \epsilon e A'_\mu J^{\mu} 
+ g_D A'_\mu \bar{\chi}\gamma^\mu \chi \,, 
\label{portal}
\en
where $g_D$ is the coupling of the dark photon with dark fermions, $e$ is the electric charge, and $J_\mu$ the electromagnetic current composed of the SM fermions. Note the coupling of the $A'_\mu$ with 
$J^{\mu}$ is obtained after the shift of electromagnetic field $A_\mu \to A_\mu + \epsilon A'_\mu$ leading to the removal of the mixing term~(\ref{Lmix}).  

The decay width of the $A'$ to the dark fermions is 
\eq 
\Gamma_{A'\to \bar{\chi}\chi } = \frac{\alpha_D}{3} 
\, m_{A'} \, 
(1 + 2 y_\chi^2) \, (1 - 4 y_\chi^2)^{1/2} \,, 
\label{AtoDark}
\en 
where  $y_\chi = m_\chi/m_{A'}$,  $\alpha_D=g_D^2/4\pi$, $m_{A'}$ and $m_\chi$ are masses of dark photon and dark fermions, respectively. 

One should note that the shift $A^\mu \to A^\mu + \epsilon A'^\mu$ leads to a possibility of invisible or semi-visible 
decays of neutral mesons those production cross section was studied before. 

\subsection{Vector meson}

We derive effective Lagrangian describing transition of neutral vector meson to 
dark photon 
\eq
{\cal L}_{V-A'} &=&  e  \epsilon g_{V}  V_\mu A'^\mu,
\label{VAdark}
\en
using Lagrangian defining the $V-\gamma$ 
coupling~\cite{Jegerlehner:2011ti,Kachanovich:2021eqa} and 
shift of electromagnetic field $A^\mu \to A^\mu + \epsilon A'^\mu$. 
Here $g_V$ is the vector meson decay coupling, which for the $\omega$ meson is $g_\omega=0.132$ GeV.  

The width of the decay of vector meson into the dark fermion pair 
$V\to \bar{\chi}\chi$ is given by 
\eq
\Gamma(V\to \bar{\chi}\chi )  = \frac{\alpha_D (\epsilon e)^2}{3} g_V^2 \frac{(m_V^2+2m_\chi^2)\sqrt{m_V^2-4m_\chi^2}}{(m^2_{A'}-m_V^2)^2
+\Gamma^2_{A'\to \bar{\chi}\chi }m^2_{A'}} \,, 
\en
where 
$m_{A'}$ and $m_\chi$ are the masses of intermediate dark 
photon and DM fermion, respectively, 
$m_V$ is the mass of vector meson. Here we use the Breit-Wigner propagator for the dark photon $A'$ assuming  that its total width is dominated by the $A' \to \bar{\chi}\chi$ mode.  

\subsection{Pseudoscalar mesons}

Using the Wess-Zumino-Witten (WZW) effective action~\cite{Wess:1971yu,Witten:1983tw} producing 
the chiral anomaly transition of the $\pi^0$ into 
two photons and couplings of quarks with $A$ and 
$A'$ we can generate the amplitudes describing 
decays $\pi^0 \to \gamma\gamma$, 
$\pi^0 \to \gamma A'$, and $\pi^0 \to A' A'$: 
\eq
A(\pi^0 \to \gamma\gamma) &=& \frac{\alpha}{\pi F_\pi}\epsilon^{\mu\nu\alpha\beta}\epsilon_{\mu}k_{1\nu}\epsilon_{\alpha}k_{2\beta}\,, 
\\
A(\pi^0 \to \gamma A') &=& \frac{\alpha \epsilon}{\pi F_\pi}\epsilon^{\mu\nu\alpha\beta}\epsilon_{\mu}k_{1\nu}\epsilon_{\alpha}k'_{2\beta}\,, 
\\
A(\pi^0 \to A' A') &=& \frac{\alpha \epsilon^2}{\pi F_\pi}\epsilon^{\mu\nu\alpha\beta}\epsilon_{\mu}k'_{1\nu}\epsilon_{\alpha}k'_{2\beta}
\en
where $\epsilon_{\mu}$ and $\epsilon_{\alpha}$ are 
polarization vectors of photon and  dark photon respectively, 
$k_1$ and $k_2$ are the momenta of the final states, 
$F_\pi = 92.4$ MeV is the pion decay constant, $\alpha = 1/137.036$ is fine structure constant.
The decay widths of the $\pi^0$ into 
$\gamma\gamma$, $\gamma A'$, and $A' A'$ are given by 
\eq 
\Gamma(\pi^0 \to \gamma\gamma)&=&\frac{\alpha^2}
{64 \pi^3} \, \frac{m_\pi^3}{F_\pi^2}
\,, \\
\Gamma(\pi^0 \to \gamma A')&=&
\frac{\alpha^2\epsilon^2}{64\pi^3} \, 
\, \frac{m_\pi^3}{F_\pi^2} \, 
\Big(1 - \frac{m^2_{A'}}{m_\pi^2}\Big)^3 \,, \\
\Gamma(\pi^0 \to A' A')&=&\frac{\alpha^2\epsilon^4}{64\pi^3} \, \frac{m_\pi^3}{F_\pi^2} \,
\Big(1 - \frac{4 m^2_{A'}}{m_\pi^2}\Big)^{3/2} \,.
\en
Here in case of the decays into two identical 
particles we take into account the combinatorial 
factor $1/2$. One can see that these decay rates obey 
the relations: 
\eq 
\Gamma(\pi^0 \to \gamma\gamma) : 
\Gamma(\pi^0 \to \gamma A')    : 
\Gamma(\pi^0 \to A' A') = 1 :  \epsilon^2 \Big(1 - \frac{m^2_{A'}}{m_\pi^2}\Big)^3 : 
\epsilon^4 
\Big(1 - \frac{4 m^2_{A'}}{m_\pi^2}\Big)^{3/2} \,.
\en 
Due to the fact that $\epsilon$ mixing parameter is small, we want to note that semi-invisible decay channel $\pi^0\to \gamma A'$ 
provides a better sensitivity for dark photon. 

Besides, we need to take into account possible decay 
of the dark photon $A'$ into dark fermion-antifermion pair  
$\gamma \chi \bar{\chi}$. It leads to a probability 
of the three-body decay $\pi^0 \to \gamma\chi\bar{\chi}$  
of the neutral pion. The corresponding decay distribution 
reads 
\eq
d\Gamma(\pi \to \gamma\chi\bar{\chi} ) = \frac{\alpha^2\epsilon^2 \alpha_D }{192 \pi^4 F_\pi^2 m_\pi^3} \, 
(m_\pi^2-q^2)^3(q^2+2m_\chi^2)(q^2-4m_\chi^2)^{\frac{1}{2}} \frac{ 1}{(m^2_{A'}-q^2)^2+\Gamma^2_{A'\to \bar{\chi}\chi }m^2_{A'}} \frac{dq^2}{\sqrt{q^2}} \,,
\en
where $q^2$ should be integrated from $4 m_\chi^2$ 
to $m_\pi^2$, the decay width 
$\Gamma_{A'\to \bar{\chi}\chi }$ 
is defined in Eq.~(\ref{AtoDark}).

In previous section, it was shown that in the case 
of pion charge-exchange scattering at a nuclear target 
one has a sizable yield contribution of the $f_2(1270)$ mesons having spin-parity $2^+$. 
For this meson the dominant decay mode is the one into two 
pions~\cite{Dobado:2001rv,Workman:2022ynf}. Therefore, we will take into account that the $f_2(1270)$ meson gives an additional yield to neutral pions. 

For an estimate of the yield of the $\eta$ and $\eta'$ mesons we should to take into account 
their mixing. Here we will follow the scheme of the 
octet-singlet mixing proposed in Refs.~\cite{Feldmann:1998vh,Feldmann:1999uf}. 
In particular, in this scheme the two-photon decay rates 
of the $\eta$ and $\eta'$ mesons 
are given by~\cite{Feldmann:1998vh,Feldmann:1999uf}  
\eq
\Gamma(\eta \to \gamma A')= 
\frac{9\alpha^2 \epsilon^2}{16\pi^3} \, m_\eta^3 \,
\Big(1 - \frac{m^2_{A'}}{m_\eta^2}\Big)^3 \
\Bigg[ \frac{C_8 \cos\theta_0}{f_8\cos(\theta_8-\theta_0)} 
- \frac{(1-\Lambda_3)C_0 \sin\theta_8}{f_0\cos (\theta_8-\theta_0)}\Bigg]^2 \,, \\
\Gamma(\eta' \to \gamma A')= \frac{9\alpha^2 \epsilon^2}{16\pi^3} \, m_{\eta'}^3 \,
\Big(1 - \frac{m^2_{A'}}{m_{\eta'}^2}\Big)^3 \ 
 \Bigg[ \frac{C_8 \sin\theta_0}{f_8\cos(\theta_8 - \theta_0)} 
+ \frac{(1-\Lambda_3)C_0 \cos\theta_8}{f_0\cos (\theta_8-\theta_0)}\Bigg]^2 \,,  
\en
where $C_8=(e_u^2+e_d^2-2e_s^2)/\sqrt{6}$ and $C_0=(e_u^2+e_d^2+e_s^2)/\sqrt{3}$ are 
the charge factors, $\theta_8 = -(21.2 \pm 1.6)^{\circ}$  and $\theta_0 = -(9.2 \pm 1.7)^{\circ}$  
are the mixing angles, $f_8 = (1.26 \pm 0.04) \sqrt{2}F_\pi$ and  $f_0 = (1.17 \pm 0.03) \sqrt{2}F_\pi$ are the octet and single leptonic decay constants, $\Lambda_3$ is 
the OZI-rule violating parameter $-0.28 < \Lambda_3 < 0.02$. 
One should stress that for the $\eta'$ meson it is more interesting to the study decay $\eta' \to \rho^0\gamma$ with next conversion of virtual $\rho^0$ meson into dark photon. 
Branching of the process $\eta' \to \rho^0\gamma$ 
is $\sim 30 \%$ of $\eta'$. 
The amplitude and decay width of this process are given respectively
by~\cite{Dumbrajs:1983jd,Feldmann:1999uf}
\eq
A(\eta'\to \rho^0\gamma) &=& 
e g_{\eta'\rho\gamma} \,\epsilon^{\mu\nu\alpha\beta}\epsilon_{\mu} p_{\nu}\epsilon_{\alpha} k_{\beta}\,, 
\\
\nonumber\\
\Gamma(\eta'\to \rho^0\gamma) &=& 
\frac{\alpha}{8 m_{\eta'}^3} g_{\eta'\rho\gamma}^2 (m_{\eta'}^2-m_\rho^2)^3  \,, 
\en
where 
$\epsilon_{\mu}$ and $\epsilon_{\alpha}$ are 
the polarization vectors of photon and $\rho^0$ 
meson respectively, $p$ and $k$ are the momenta of the 
$\eta'$ and $\rho^0$ meson, 
$g_{\eta'\rho\gamma}=1.257$ GeV$^{-1}$ 
is the effective $\eta'\rho\gamma$ coupling fixed from data~\cite{Workman:2022ynf}. This value is 
in good agreement with the theoretical prediction done 
in Refs.~\cite{Feldmann:1998vh,Feldmann:1999uf}.

Using kinetic mixing Lagrangian, we can receive two different decays. The first  one is decay of $\eta'$ meson into $\rho^0$ meson and dark photon $A'$    
\eq
\Gamma(\eta'\to \rho^0 A') &=& 
\frac{\alpha \epsilon^2 g_{\eta'\rho\gamma}^2}{8 m_{\eta'}^3}  \, 
\lambda^{\frac{3}{2}}(m^2_{\eta'},m^2_\rho,m^2_{A'}) \,, 
\en
and with next decay where dark photon transits to dark fermions
\eq
\Gamma(\eta'\to \rho^0 \bar{\chi}\chi)&=& \frac{\alpha\alpha_D\epsilon^2 g^2_{\eta'\rho\gamma}}{24\pi m^3_{\eta'}} \, 
\lambda^{\frac{3}{2}}(m^2_{\eta'},m^2_\rho,q^2) \, (q^2-4m^2_{\chi})^{\frac{1}{2}} \, \Big[q^2+2m^2_\chi \Big] \nn\\
 &\times& \frac{ 1}{(m^2_{A'}-q^2)^2+\Gamma^2_{A'\to \bar{\chi}\chi }m^2_{A'}} \frac{dq^2}{\sqrt{q^2}}
\,,
\label{PS2Rho2Chi}
\en
where  $\lambda(x, y, z) = x^2 + y^2 + z^2 - 2xy - 2xz - 2yz$ 
is the K\"allen kinematical triangle function. 
Needed to note that these semi-invisible decays do not have additional $\alpha$ suppression factor which we have in case of decays $\eta(\eta') \to \gamma A'$ .

The second decay is process with intermediate 
conversion of the $\rho^0$ mesons 
into dark photon is given by the expression 
\eq
\Gamma(\eta'\to (\rho^0\to A')\gamma)= \frac{\alpha\epsilon^2}{8 m_{\eta'}^3} g_{\eta'\rho\gamma}^2 (m_{\eta'}^2-m_{A'}^2)^3 \frac{g_\rho^2}{(m^2_{\rho}-m_{A'}^2)^2+\Gamma^2_{\rho}m^2_{\rho}}.
\en
The differential decay width of $\eta' \to \gamma \chi\bar{\chi}$ with $\rho$ 
meson resonance transition is 
 \eq
d\Gamma(\eta'\to (\rho^0\to A'\to \chi\bar{\chi})\gamma)&=&
\frac{\alpha^2\epsilon^2\alpha_D }{ 6 m_{\eta'}^3}  (q^2-m_{\eta'}^2)^3 (q^2+2m_\chi^2)(q^2-4m_\chi^2)^{\frac{1}{2}}  \nn
\\
&\times&\Bigg[
\frac{g_\rho^2}{(m^2_{A'}-q^2)^2
+\Gamma^2_{A'\to \bar{\chi}\chi }m^2_{A'}} 
\, \frac{g_{\eta'\rho\gamma}^2}{(m^2_{\rho}-q^2)^2+\Gamma^2_{\rho}m^2_{\rho}}
\Bigg]\frac{dq^2}{\sqrt{q^2}}.
\label{WithRhoDecayPS}
\en
Where important to note that $\eta$ meson also can decay to $\bar{\chi}\chi \gamma$ through intermediate $\rho^0$ meson which transit to dark fermions by mixing with dark photon (see Lagrangian in eq.(\ref{VAdark})). The coupling of $g_{\eta\rho\gamma}=1.52$ GeV$^{-1}$ in framework of FKS scheme and $1.42$ GeV$^{-1}$ from experimental data \cite{Feldmann:1998vh,Feldmann:1999uf}. Wherein from analysis of decay $\eta \to \gamma\pi\pi$ is known that contribution due to intermediate $\rho^0$ vector meson is huge, for decay of $\eta' \to \gamma\pi\pi$ is dominate \cite{Venugopal:1998fq,Benayoun:2003we}. In case of decay to dark fermions we have more large area of integration for light masses of dark fermions. Because from these decay we should obtain a stricter restriction than we have from process of decay pseudoscalar meson into two photon by the chiral anomaly \cite{Wess:1971yu} with one photon mixed with dark photon.   

Besides, in analogy with contribution $f_2(1270)$ to neutral pion yield, we can receive additional yield of $\pi^0$ and $\eta$ mesons from decays $\eta \to 3\pi^0$, 
$\eta'\to \pi^0\pi^0\eta$, and $\eta'\to \pi^+\pi^-\eta$ with very sizable branchings: 
${\rm Br}(\eta \to 3\pi^0)=32.57\%$, 
${\rm Br}(\eta'\to \pi^0\pi^0\eta)=22.4\%$, 
and ${\rm Br}(\eta'\to \pi^+\pi^-\eta)=42.5\%$. For full analysis, we take into account both channel's transition to DM for $\eta'$ meson where transition through $\rho^0$ meson is the main contribution.

\begin{center}
	\begin{table*}[t]
		\centering
		\caption{Yields $N_{M^0}$ of mesons in the reaction of charge exchange at the scattering of negative pion beam onto the iron target at the beam energy of~$50$ GeV. The meson yields are based on the average value of charge exchange cross sections presented in Table \ref{tab:2}.}
		\begin{tabular}[b]{l|ccccc}
			\hline
			\hline
			 & $N_{\pi^0}$  & $N_{\eta}$ & $N_{\eta'}$ & $N_{\omega}$&$N_{f_2}$\\
            \hline
            &&&&&\\
    	NA64$_h$ ($3 \times 10^9$ \, $\pi$OT) \quad & $0.35 \times 10^6$ \quad   &  $0.089 \times 10^6$ \quad  & $0.054  \times 10^6$ \quad & $0.088  \times 10^6$ \quad & $0.089  \times 10^6$ \\
           NA64$_h$ ($5 \times 10^{12}$ \, $\pi$OT) \quad & $5.86 \times 10^8$ &  $1.48 \times 10^8$ & $0.9 \times 10^8$ & $1.47 \times 10^8$& $1.48 \times 10^8$ \\
			\hline
                \hline
			\label{ParamTable}
		\end{tabular}
  \label{yields}
	\end{table*}%
\end{center}

\section{Bounds}
\label{bounds_sec}

In this section we discuss possible bounds for the dark photon portal, which can be obtained by searching for missing energy/momentum signals in the proposed experiment NA64$_h$  with the negatively charged pion beam. Using cross section for the production of neutral mesons 
in the charge-exchange process as a result of scattering of the negative pion beam on the Fe target with 
the beam momentum $P=50$ GeV we can estimate yields of neutral mesons. The estimated yields of mesons are presented in Table~\ref{yields}. For 90\% C.L. of parameter space bounds  we will use formulas $\mathrm{Br}(M^0\to \text{inv.})\leq 2.3/N_{M^0}$ for invisible and for 
semi-invisible is $\mathrm{Br}(M^0\to \text{semi-inv.})\leq 2.3/N_{M^0}$ limits, where $N_{M^0}$ is yields of mesons. The dark portal with dark fermions includes four parameters:  $\epsilon$ kinematic mixing parameter of photon and dark photon,  $m_{A'}$ is the mass of dark photon, $\alpha_D=g_D^2/4\pi$ coupling interaction with dark fermions, and  $m_\chi$ is the dark fermion mass. These parameters can be combined into  dimensionless parameter 
$y=\alpha_D\epsilon^2(m_\chi/m_{A'})^4$ \cite{Berlin:2018bsc,Berlin:2020uwy}, which is convenient 
to use for the thermal target DM parameter space. In particular, using this dimensionless 
parameter $y=\alpha_D\epsilon^2(m_\chi/m_{A'})^4$, we 
can compare the existing and projected limits from the NA64$_h$ proposal to dark photon portal 
with dark fermions with the typical relic DM parameter space.

\begin{figure}[b]
	\includegraphics[width=0.32\textwidth,clip]{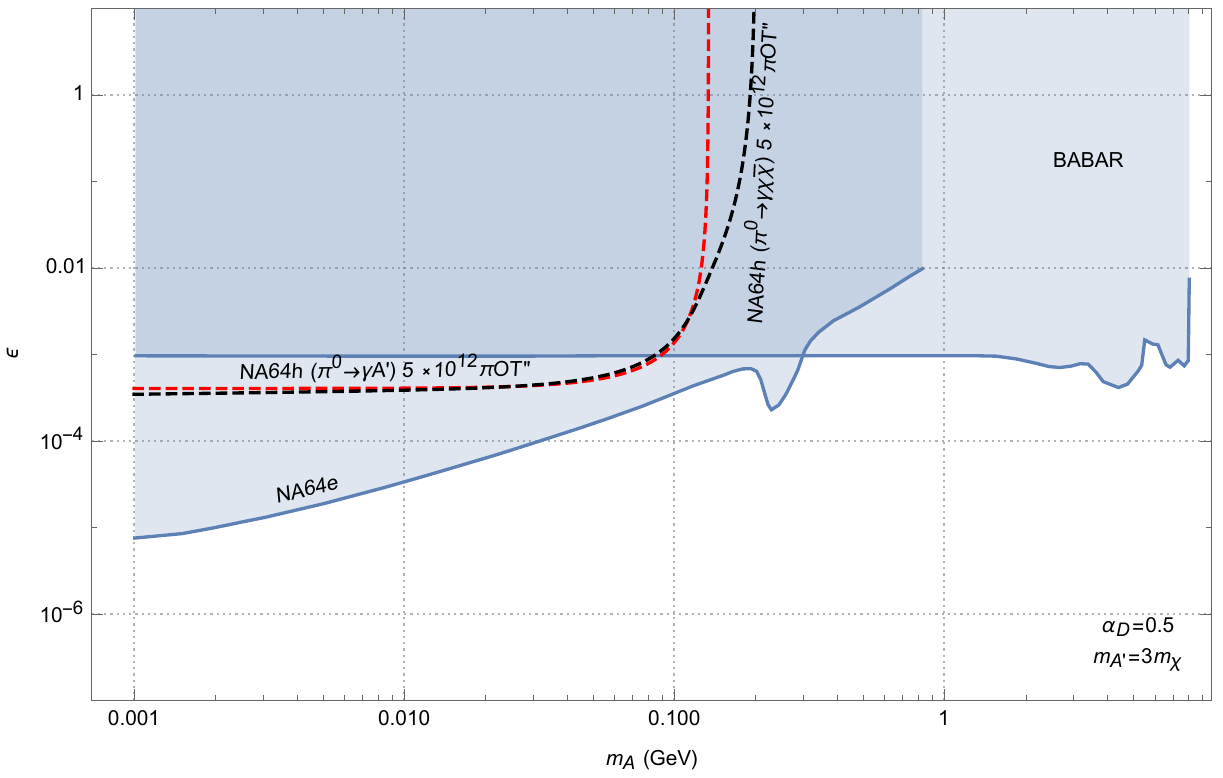}
 	\includegraphics[width=0.32\textwidth,clip]{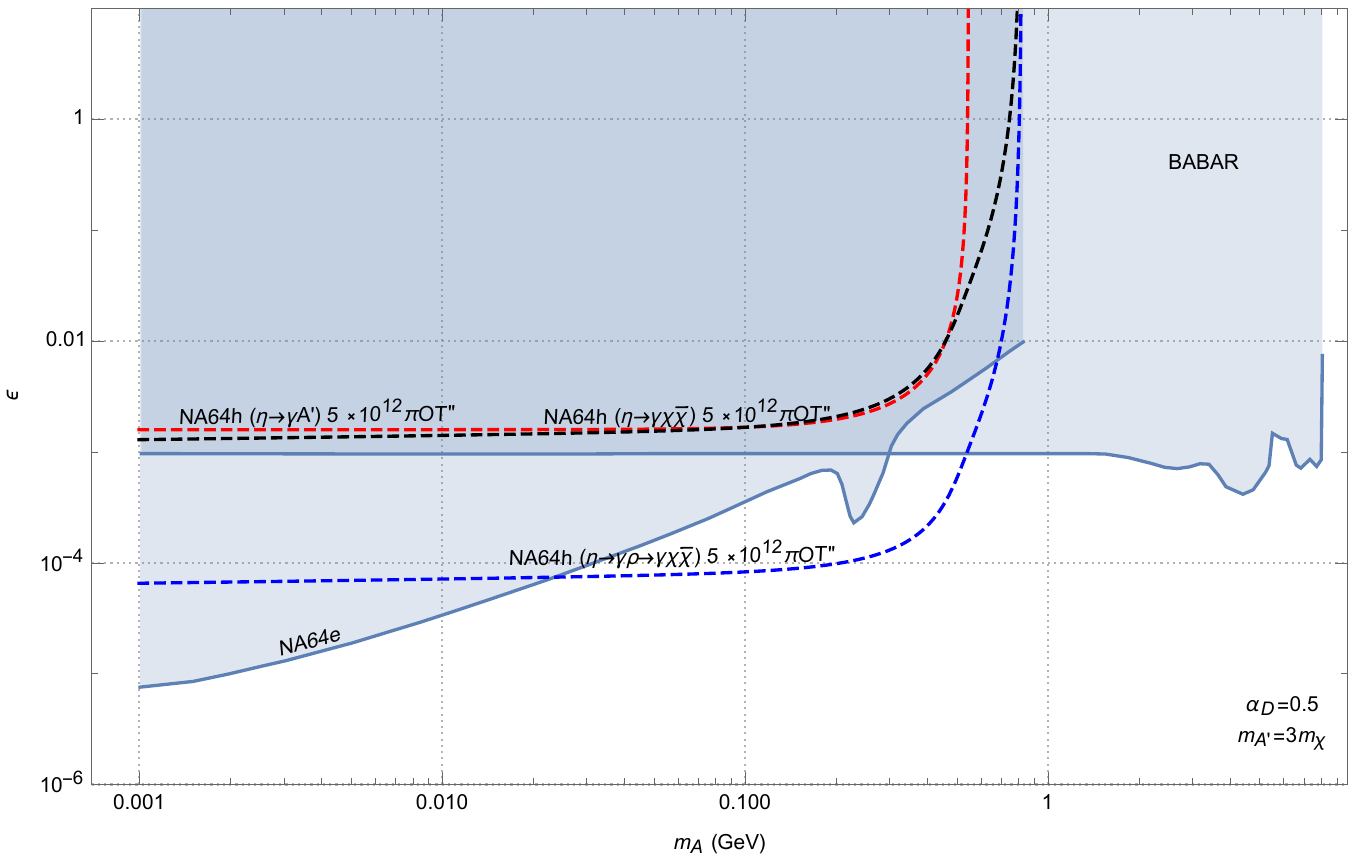}
        \includegraphics[width=0.32\textwidth,clip]{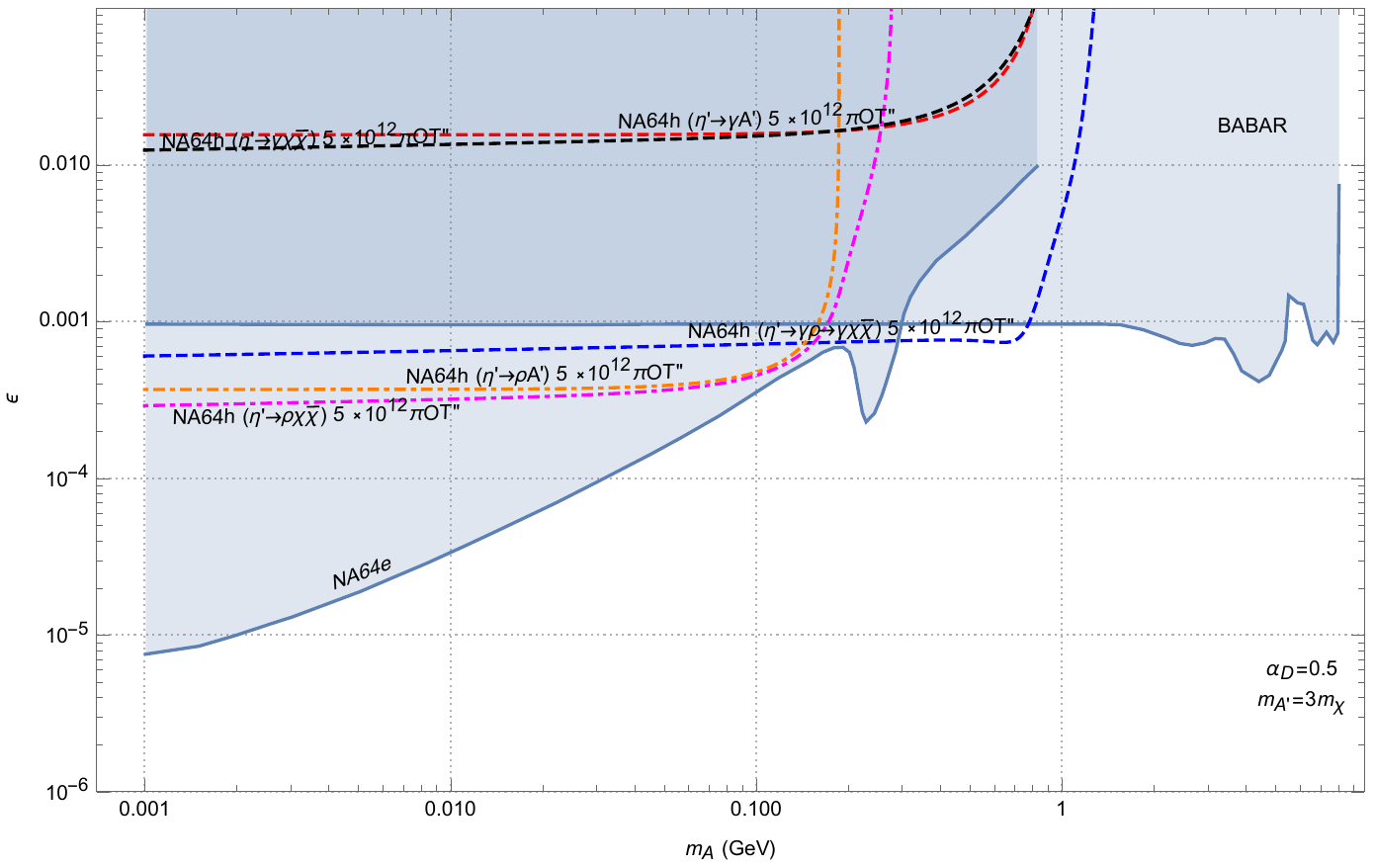}
	\caption{ Bounds for dark photon $\epsilon$ parameter mixing obtained for 90\% C.L.  and for case where $m_{A'}=3m_\chi$ and $\alpha_D=0.5$. At these panels, we show existed limit from current data of NA64$_e$ experiment~\cite{NA64:2023wbi} and constraints from production of DM in $e^+e^-$ collision at BABAR~\cite{BaBar:2013npw}. In left panel we plot the bounds from semi-invisible mode of neutral pion decays $\pi^0\to\gamma A'$ and $\pi^0\to\gamma \bar{\chi}\chi$. In central panel we  depict the bounds from semi-invisible mode of $\eta$ meson decays $\eta\to\gamma A'$, $\eta\to\gamma \bar{\chi}\chi$ and $\eta\to\gamma\rho^*\to\gamma \bar{\chi}\chi$. In right panel we show the bounds from semi-invisible mode of $\eta^\prime$ meson decays $\eta'\to\gamma A'$, $\eta'\to\gamma \bar{\chi}\chi$, $\eta'\to\gamma(\rho^{0}\to A')\to\gamma \bar{\chi}\chi$ and $\eta'\to\rho^0 A'$, $\eta'\to\rho^0 \bar{\chi}\chi$. All constrains are presented for proposal statistics $5\times 10^{12}$ $\pi$OT of NA64 experiment.} 
	\label{epsilonPS}
\end{figure}

In the previous Section it was mentioned that the real yield of neutral pions will be larger. It implies  that 
$\eta$, $\eta'$, and $f_2(1270)$ have dominant hadronic decays in which the neutral pion is is one of the possible final states. For an estimate of bounds on dark photon parameter space from pion decay we will use yield equal 
$ N_{\pi^0}= 1.14 \times 10^9$ for statistics of $5 \times 10^{12}$ $\pi$OT. We obtain factor two for the yield of $\pi^0$ from Table \ref{yields}. Full yield of $\eta$ mesons is changed insignificantly in comparison with data from Table \ref{yields} and will be equal to $N_\eta=1.9 \times 10^8$ of $\eta$ mesons for 
 the NA64 experiment with statistics in $5 \times 10^{12}$ $\pi$OT.

\begin{figure}[t]
	\includegraphics[width=0.49\textwidth,clip]{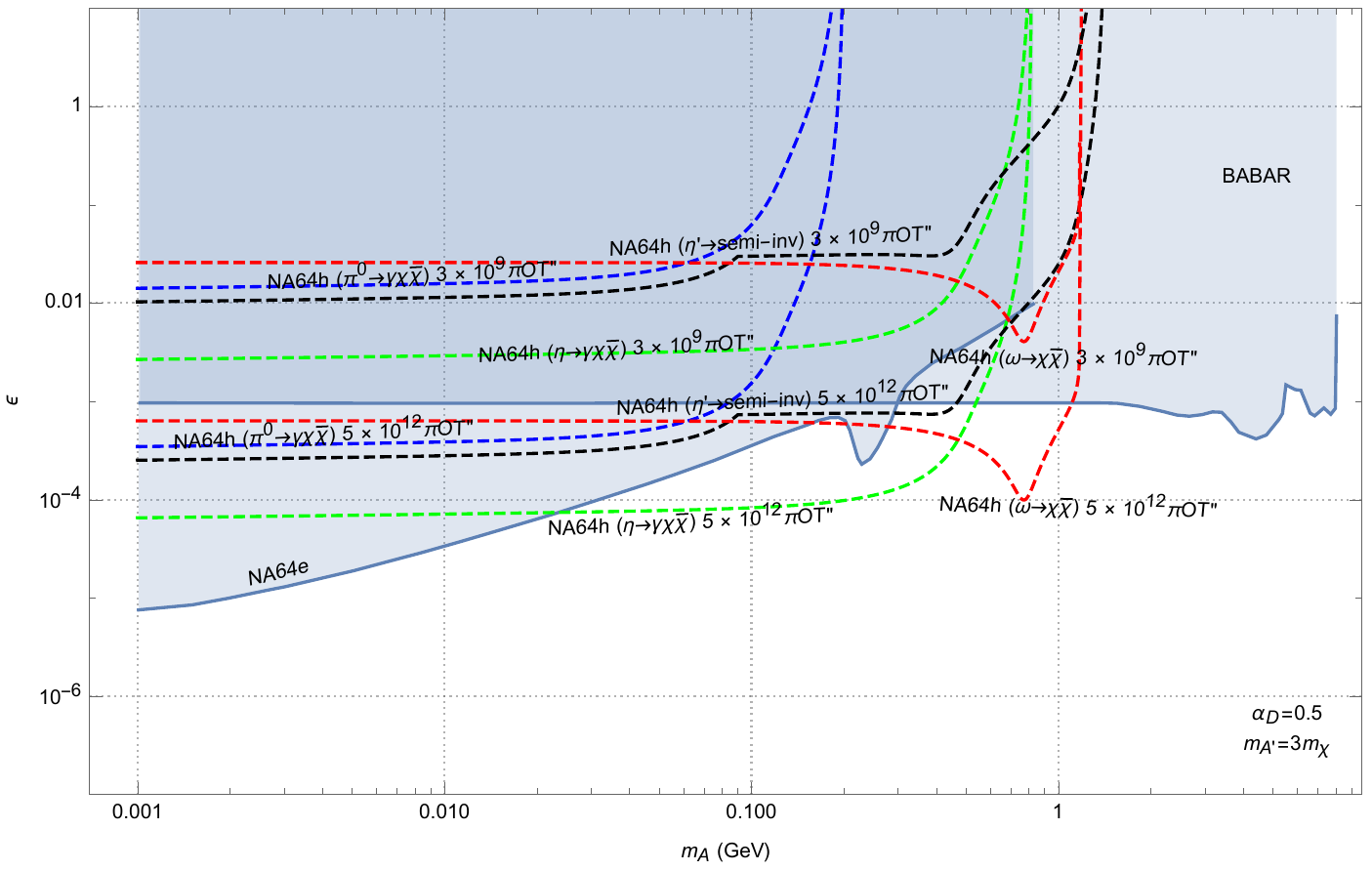}
 	\includegraphics[width=0.49\textwidth,clip]{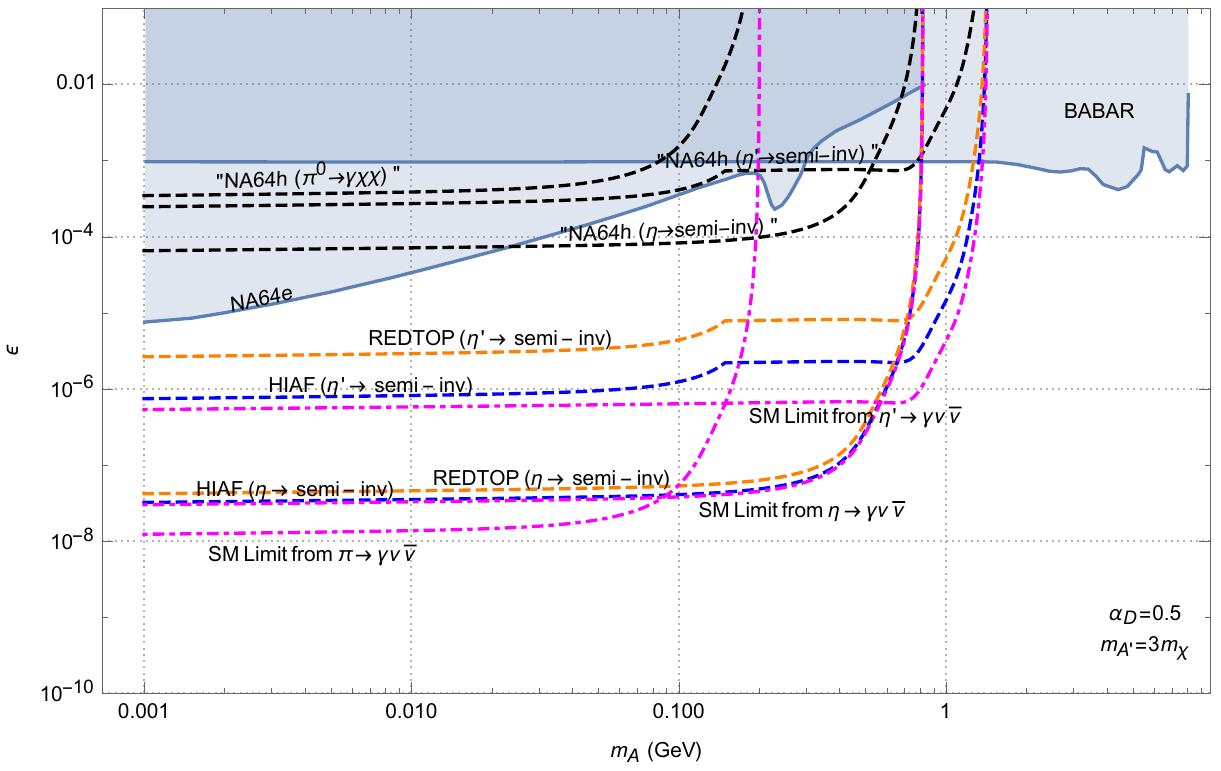}
	\caption{ Bounds for dark photon $\epsilon$ parameter mixing obtained for 90\% C.L.  and for case where $m_{A'}=3m_\chi$ and $\alpha_D=0.5$. At both panels, we show existed limit from current data of NA64$_e$ experiment~\cite{NA64:2023wbi} and constraints from production of DM in $e^+e^-$ collision at BABAR~\cite{BaBar:2013npw}.
 Left panel: Bounds from semi-invisible pseudoscalar decays ($\pi^0\to \gamma\chi\bar{\chi}$, $\eta\to \gamma\chi\bar{\chi}$, $\eta'\to semi-inv$) and invisible decay ($\omega\to \chi \bar{\chi}$) for statistics $3\times 10^9$ $\pi$OT (few days of data taking) and for $5\times 10^{12}$ $\pi$OT as proposal statistics for NA64$_h$ experiment. 
 Right panel: Bounds from semi-invisible pseudoscalar decays ($\pi^0\to \gamma\chi\bar{\chi}$, $\eta\to \gamma\chi\bar{\chi}$, $\eta'\to \gamma\chi\bar{\chi}$) for proposal/projected statistics of NA64$_h$, REDTOP~\cite{REDTOP:2022slw} and HIAF~\cite{HIAF} experiments. The dot-dashed lines show limit of neutrino floor from decay light pseudoscalar mesons to $\gamma \nu\bar{\nu}$  predicted in the framework of SM~\cite{Arnellos:1981bk}.
 }
	\label{epsilon}
\end{figure}

The analysis of limits from semi-invisible mode of pseudoscalar meson decays for dark photon parameter space ($\epsilon$ and $m_{A'}$) is shown in Fig.~\ref{epsilonPS}. We analyzed all possible channels of semi-invisible decay and obtained results that decays with finite or intermediate $\rho^0$ meson give more strict limits to parameter $\epsilon$ of kinetic mixing of dark photon and SM photon. Limits from the famous decay which is due to WZW chiral anomaly of pseudoscalar mesons into two photons and take into account that one photon mix with dark photon which also were presented in Ref.\cite{Kahn:2014sra} are suppressed as $\propto \alpha^2\epsilon^2$. Wherein decays with finite or intermediate $\rho^0$ meson are $\propto \alpha g^2_{\eta\rho\gamma}\epsilon^2$ or $\propto \alpha^2 g^2_{\eta\rho\gamma}\epsilon^2$. These factors coupled with the Breit-Wigner form of propagator produce less suppression factor for branching ratio for $\eta$ and $\eta'$ mesons. We will use notation $\eta \to \gamma\bar{\chi}\chi$ for limit from branching ratio obtained from eq.(\ref{WithRhoDecayPS}), for $\eta'$ meson we will use notation $\eta'\to semi-inv $ and use limit from branching ratio obtained from eq.(\ref{PS2Rho2Chi}) and eq.(\ref{WithRhoDecayPS}).
\begin{figure}[b]
	\includegraphics[width=0.49\textwidth,clip]{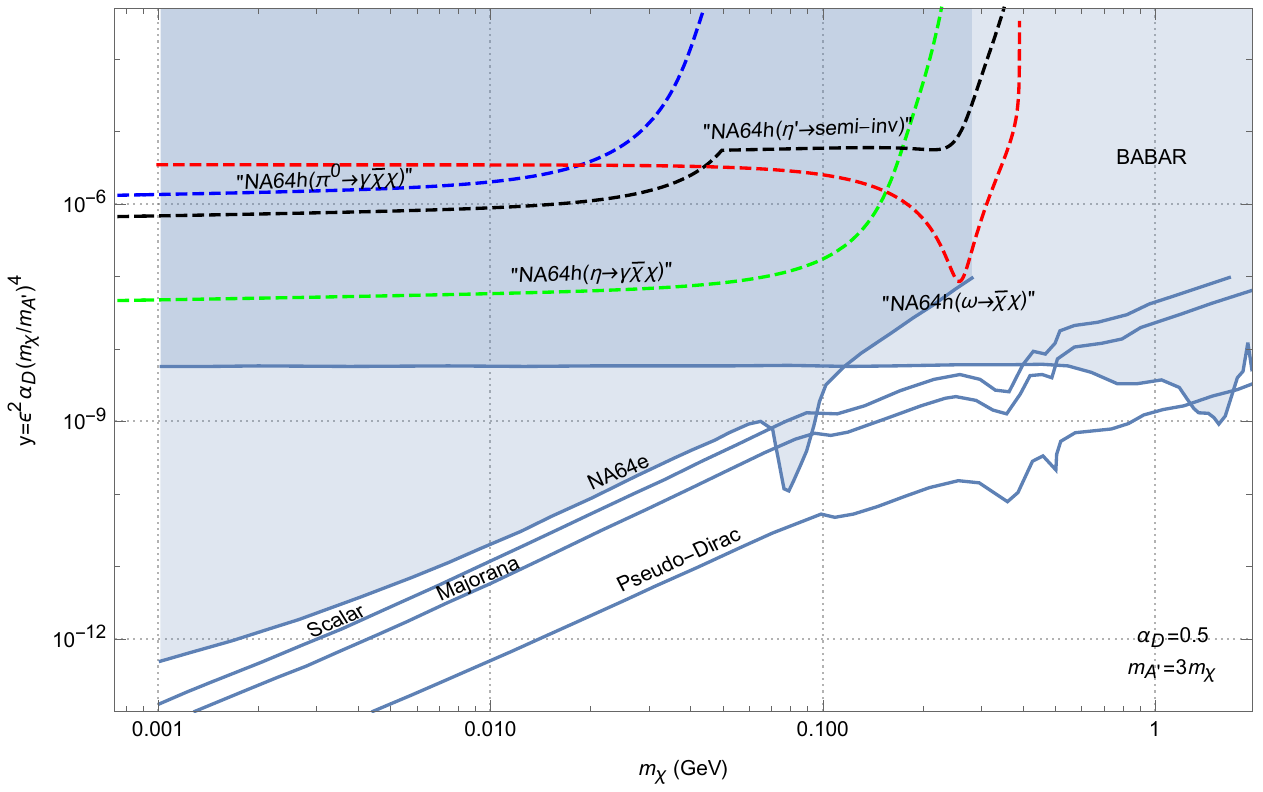}
 	\includegraphics[width=0.49\textwidth,clip]{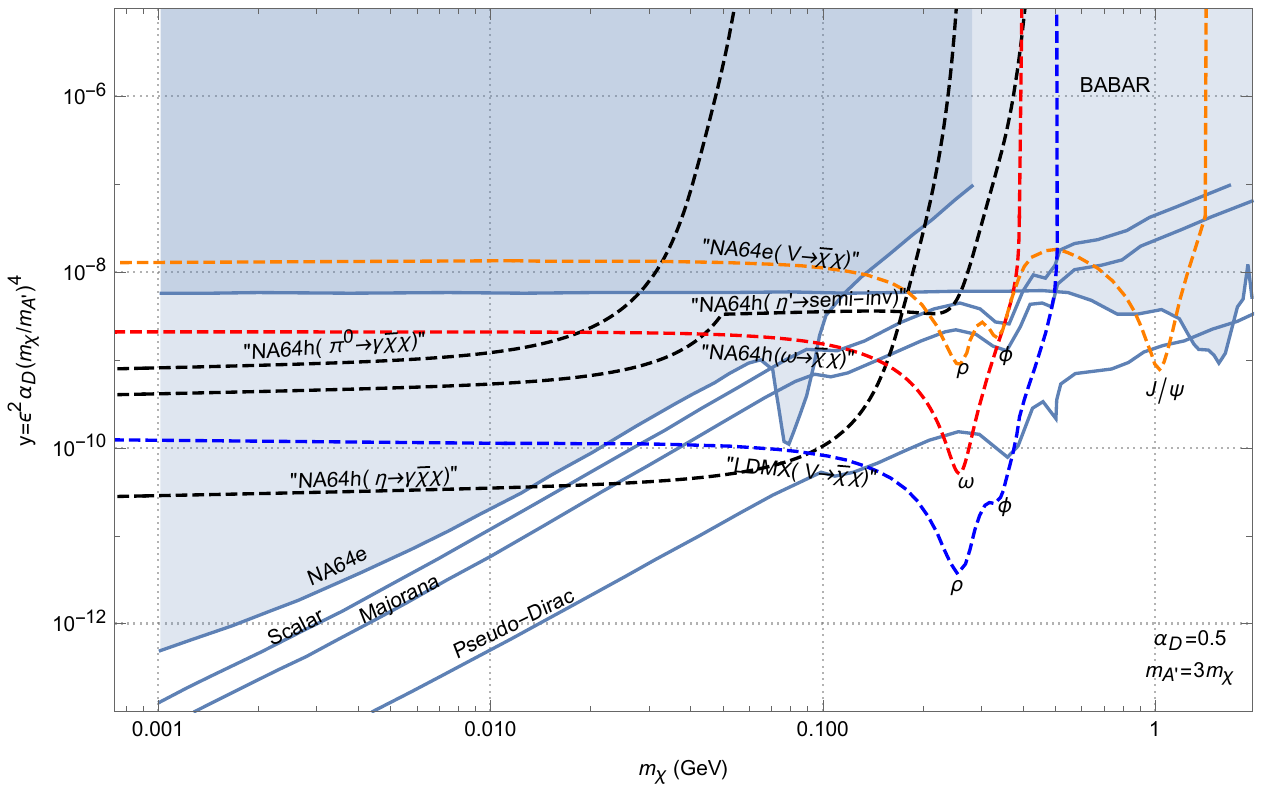}
	\caption{ The projected 90\% C.L. exclusion limits for benchmark scenarios of invisible and semi-invisible neutral meson decays to dark matter in dark photon mediator portal model. The constrains are shown for the benchmark values  $m_{A'}=3m_\chi$ and $\alpha_D=0.5$.
 At both panels, we show existed limit from last data of NA64$_e$ experiment~\cite{NA64:2023wbi} and 
 constraints from production of DM in $e^+e^-$ collision at BABAR \cite{BaBar:2013npw}. Left panel: limits from semi-invisible pseudoscalar decays ($\pi^0\to \gamma\chi\bar{\chi}$, $\eta\to \gamma\chi\bar{\chi}$, $\eta'\to semi-inv$) and invisible decay ($\omega\to \chi \bar{\chi}$) using statistics $3\times 10^9$ $\pi$OT. Right panel: Limits for the same decays for $5\times 10^{12}$ $\pi$OT of NA64$_h$ experiment in comparison with bound from invisible vector meson decays obtained for NA64$_e$ and LDMX experiments \cite{Schuster:2021mlr}.}
	\label{Yestimation}
\end{figure}

The limits for dark photon parameter space ($\epsilon$ and $m_{A'}$) using yields of neutral meson presented in Table \ref{yields} are shown in Fig.~\ref{epsilon}. 
In particular, in left panel it is shown the bound from semi-invisible pseudoscalar mesons decay to 
semi-invisible mode and for invisible decay ($\omega\to \chi \bar{\chi}$) through the transition in the dark photon by mixing with the ordinary photon. In derivation of the constraint, we use the benchmark values 
$m_{A'}=3m_\chi$ and $\alpha_D=0.5$. 
For low statistics $3\times 10^9$ $\pi$OT for the NA64$_h$ 
experiment one can see 
that all neutral meson decays cannot provide sizable bounds in comparison with existing
limits to mixing coupling from NA64$_e$ and BABAR experiments. For proposal statistics $5\times 10^{12}$ $\pi$OT we can test new area of parameter space dark photon model by study invisible and semi-invisible modes of $\eta$, $\eta'$ and $\omega$ mesons.  The strong
limits from vector meson transition to  dark photon is connected with decay width $\omega\to \chi \bar{\chi}$ that is  proportional only to 
$\propto\alpha\alpha_D\epsilon^2$.
Besides for $\eta'$ mesons,  the dominant contribution to the 
decay width $semi-inv$ is absent for the decay $\eta'\to\gamma\rho^0$ process. The decay  
$\eta'\to semi-inv$ with $\rho^0$ transition is shown by the black dashed line. For $\eta$ meson limit for parameter mixing $\epsilon$ also is strong and can test new area of parameter space ($\epsilon$ and $m_{A'}$). With statistics of  $5\times 10^{12}$ $\pi$OT,  the NA64$_h$ can probe the parameter space for the kinetic mixing of dark photons which is currently unexplored.  
In the right panel in Fig.~\ref{epsilon} it is seen that limits from projected $\eta$/$\eta'$ factories 
REDTOP~\cite{REDTOP:2022slw} (with projected yields $\sim 3.9 \times 10^{14}$ for 
$\eta$ and $\sim 7.9 \times 10^{11}$ for $\eta'$) and HIAF (with projected yields $\sim 10^{15}$ for $\eta$ and 
$\sim 10^{13}$ for $\eta'$) \cite{HIAF} can constrain the mixing parameter $\epsilon$ at level $10^{-6}$ and $10^{-7}$  for the $A'$ mass range up to masses of $\eta/\eta'$ mesons. Herewith we need to note that the neutrino 
floor limit~\cite{Arnellos:1981bk} for this process is sufficiently strong 
and the main signal of missing energy should be associated with the
transition to dark sector. In Fig.~\ref{epsilon} the regarding  bounds are shown  by dot-dashed lines. For $\eta$ meson neutrino  floor~\cite{Arnellos:1981bk} is very close to the projected sensitivity  of future  $\eta$/$\eta'$ factories.

Using projected statistics for the neutral meson yield in the NA64, we can predict the typical bound on the semi-invisible on the invisible branching  ratio
\eq
\mbox{Br}(\pi^0\to\gamma +A') &<& 1.58 \times 10^{-9} \,\, \quad (\text{from NA64$_h$}); \nn\\
\mbox{Br}(\eta\to semi-inv) &<& 9.4 \times 10^{-9}  \,\,\,\,\,\quad(\text{from NA64$_h$}); \nn\\
\mbox{Br}(\eta'\to semi-inv) &<& 4.7 \times 10^{-9}  \,\,\,\,\,\quad(\text{from NA64$_h$}); \\
\mbox{Br}(\omega\to \mbox{inv.}) &<& 1.63 \times 10^{-10}     \quad (\text{from NA64$_h$}); \nn  
\en
in the framework of 90\% C.L. of missing energy signature implying zero signal events  and background free 
case. This branching is calculated for case if $A'$ dark photon decay to dark fermions with $\alpha_D=0.5$ and 
$m_{A'}=3m_\chi$. The proposed REDTOP~\cite{REDTOP:2022slw} and HIAF~\cite{HIAF} factories are expected to work at the level of neutrino floor, $\mbox{Br}(\eta\to \gamma\nu\bar{\nu})\simeq 2\times 10^{-15}$,  for $\eta'$.  
(see right panel in Fig.~\ref{epsilon}). The  existing limit for this branching ratio of neutral pion from the NA62 
experiment is $\lesssim 1.9\times 10^{-7}$. This limit was obtained by NA62 with 
$N_{\pi^0}=4.12 \times 10^8$ $\pi$OT~\cite{NA62:2019meo}. In the framework of SM decay of pion into photon and
pair of neutrinos is $\mbox{Br}(\pi^0\to \gamma\nu\bar{\nu})\simeq 2 \times 10^{-18}$~\cite{Arnellos:1981bk}.

In Fig.~\ref{Yestimation} we present the constraints on the dimensionless parameter $y=\alpha_D\epsilon^2(m_\chi/m_{A'})^4$ and the mass of dark fermions 
$m_\chi$~\cite{Berlin:2018bsc}. For the existing and projected limits  of NA64$_h$ one can compare  this parameter  with the typical relic DM parameter space.  The yields of 
pseudoscalar mesons are too small to test unconstrained 
area for dark photon model if we consider decay from WZW term with taking into account mixing with dark photon. Including decay of $\eta$ and $\eta'$ mesons with intermediate or finite state $\rho^0$ vector meson, we can test and constrain  new area for dark photon model.
Besides, vector meson invisible decay can give a relatively strong bound to the dark photon model with dark 
fermions~\cite{Schuster:2021mlr,Zhevlakov:2023wel}.
For the case of a negative pion beam we predict a sufficiently large  yield of $\omega$ vector bosons (see Table.\ref{yields}). 
The NA64$_h$ limit from $\omega$ in charge exchange reaction for statistics $5\times 10^{12}$ 
$\pi$OT will give a limit between the projected bounds from invisible meson decay limits computed for the electron beam experiments NA64$_e$ with  $5\times 10^{12}$ electron on target (EOT) and projected bounds LDMX 
with $10^{16}$~EOT~\cite{Schuster:2021mlr}, respectively. 
The process of the $\omega$ production in charge exchange reaction and $\rho^0$ meson production presented 
in Ref.~\cite{Zhevlakov:2023wel} gives relatively weak bounds but in the case of invisible decays it is associated with the same signature of missing energy. 
Pseudoscalar $\eta$  meson will test area near of peak from analysis NA64 experiment with positron beam \cite{NA64:2023ehh}. Information from semi-invisible $\eta$ meson decay and $\omega$ are covered possible limits which can be obtained from $\eta'$ meson study.

\begin{figure}[t]
	\includegraphics[width=0.49\textwidth,clip]{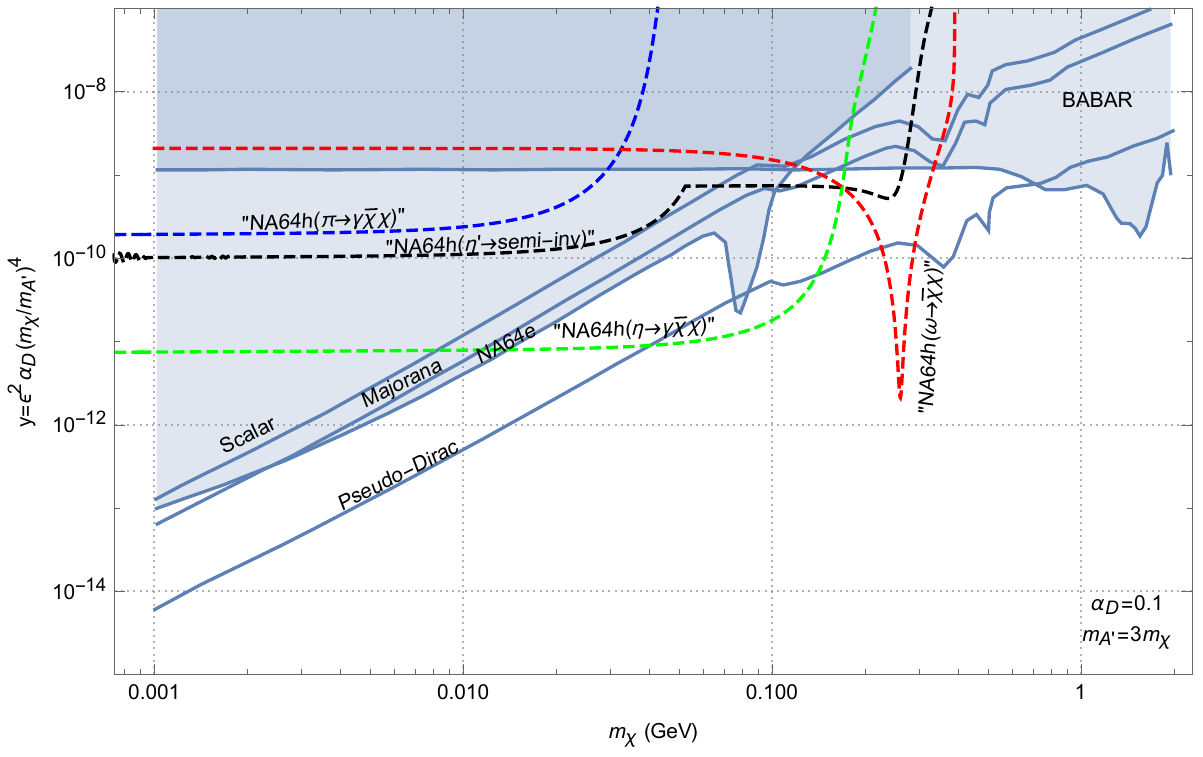}
 	\includegraphics[width=0.49\textwidth,clip]{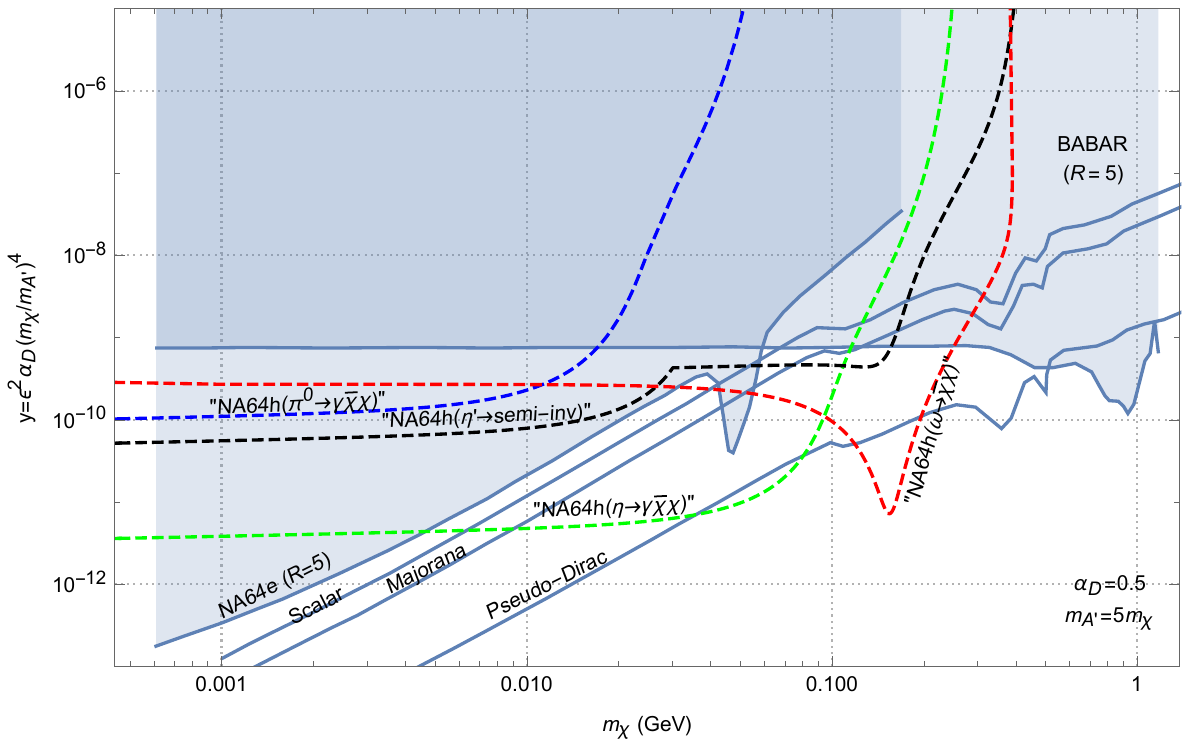}
	\caption{The projected 90\% C.L. exclusion for benchmark scenarios of invisible and semi-invisible neutral meson decays to dark matter in dark photon mediator portal model. At both panels, we show existed limit from last data of NA64$_e$ experiment~\cite{NA64:2023wbi} and constraints from production of DM in $e^+e^-$ collision at BABAR \cite{BaBar:2013npw} and limits from invisible pseudoscalar decays ($\pi^0\to \gamma\chi\bar{\chi}$, $\eta\to \gamma\chi\bar{\chi}$, $\eta'\to semi-inv$) and invisible decay ($\omega\to \chi \bar{\chi}$) using statistics $5\times 10^{12}$ $\pi$OT. Left panel: for parameter $\alpha_D=0.1$ and $m_{A'}=3m_\chi$. Right panel: for parameter $\alpha_D=0.5$ and $m_{A'}=5m_\chi$. }
	\label{DiagdiffR_alpha}
\end{figure}

In Fig.\ref{DiagdiffR_alpha} we show typical parameter space associated with various benchmark values of dark photon 
parameters. The variations of existing constraints from  data of the NA64$_e$ experiment~\cite{NA64:2023wbi} and  from 
production of DM in $e^+e^-$ collision at BABAR \cite{BaBar:2013npw} were done based on analysis of 
$R=m_{A'}/m_\chi$ dependence presented in Ref.~\cite{Berlin:2020uwy}. Changing the typical values of $\alpha_D$  affects the constraints of vector meson only in the area near the mass of vector mesons. 
 Pseudoscalar meson constraints are shifted with
the  existing constraints from the Bremsstrahlung process at NA64$_e$ 
experiment.  Relatively small values of $\alpha_D$ lead to the suppressed branching ratio for semi-invisible 
decay of pseudoscalar mesons. Note that the dependence on mass ration $R=m_{A'}/m_\chi$ is  crucial for all channels.  For
$R \gtrsim 5$ the  pseudoscalar semi-invisible decay can rule out a new region of the parameter space of dark photon model with  dark fermions. 


\section{conclusion}
\label{Conclusion}
The yield of neutral mesons associated with the charge exchange process was calculated in the 
framework of model-independent Regge approach using previous experimental data for charge exchange reactions. For mesons with masses $m_{M^0} > m_{\pi^0}$  the  current precision for  their  total production cross section 
is at the level $\simeq 30-35\%$.  This will constraint the sensitivity of the future searches for neutral meson decays into the invisible or the semi-invisible modes, and thus, 
this precision should be improved by at least a factor of $\gtrsim$ 3. 

Based on averaged yields of neutral mesons in charge exchange reactions, we made an estimate of semi-invisible decays of light pseudoscalar mesons and invisible decay of vector 
$\omega$ meson. The obtained results for the parameter space of the dark photon model with dark fermions show that vector meson production in negative pion beam scattering on a fixed target has 
an advantage in comparison with the leptonic beam. We also shown that $\eta$ and $\eta'$ mesons can be used for probing dark vector portal. Besides, the studies of meson decay with sufficiently large statistics provide an 
opportunity to test dark  vector portal for unconstrained area of the  parameter space of the scenario. 
In addition  the hadronic beam opens a new possibility to test dark 
matter physics by analyzing rare invisible or semi-invisible decays of mesons by using the missing 
energy/momentum technique. 

Also we would like to note that this calculation is associated with the charge exchange process. However, 
the realistic number  of neutral pions with small recoil energy to the target can be larger in the process of pion scattering on the atomic target. In particular, neutral meson production can be also associated with two-meson 
production in the  double-Regge region~\cite{JPAC:2021rxu} or by the Primakoff effect from Bremsstrahlung 
photons.  The full picture requires simulations in GEANT4~\cite{Allison:2006ve}. 
Needed to note that neutral vector meson decay to invisible mode to DM is more convenient for analysis dark photon portal 
model because in this case all the energy of vector meson will be missed. For semi-invisible decays of neutral pseudoscalar 
meson we have another kinematic picture. 

Besides, we need to note that the experimental study of DM physics by searching for invisible or semi-invisible meson 
decays with the missing energy technique needs to deepen our knowledge in meson production physics with the hadronic and 
leptonic beam at high energy. Model-independent approaches that are based on experimental data are required.

\begin{acknowledgments}

We thank Xurong Chen for discussion on projected characteristics of $\eta$ and $\eta'$ mesons factory at future HIAF facility and Nikolai Krasnikov for valuable discussions. This work was funded by ANID PIA/APOYO AFB180002 and 
AFB220004 (Chile), by FONDECYT (Chile) under Grant No. 1230160 
and by ANID$-$Millen\-nium Program$-$ICN2019\_044 (Chile).
The work of A.~S.~Zh. on exclusion limits calculation 
is supported by Russian Science Foundation (grant No. RSF 23-22-00041). 
The work of A.~S.~Zh. on analyze of rare decays is supported by the Foundation 
for the Advancement of Theoretical Physics and Mathematics "BASIS".

\end{acknowledgments}

\bibliography{bibPiRecharge.bib}
  
\end{document}